\documentclass[12pt]{article}

\usepackage{amsmath}
\usepackage{amssymb}
\usepackage{epsfig}

\newcommand{\capdef}{}
\newcommand{\mycaption}[2][\capdef]{\renewcommand{\capdef}{#2}%
        \caption[#1]{{\itshape #2}}} 
\makeatletter
\renewcommand{\fnum@table}{\textbf{\tablename~\thetable}}
\renewcommand{\fnum@figure}{\textbf{\figurename~\thefigure}}
\makeatother

\def\ltap{\ \raisebox{-.4ex}{\rlap{$\sim$}} \raisebox{.4ex}{$<$}\ }
\def\gtap{\ \raisebox{-.4ex}{\rlap{$\sim$}} \raisebox{.4ex}{$>$}\ }

\newcounter{myenumi}

\renewcommand{\themyenumi}{\roman{myenumi}}
{\end{list}}

\newlength{\myem}
\newcommand{\sep}[1]{#1}
\newcounter{mysubequation}[equation]
\renewcommand{\themysubequation}{\alph{mysubequation}}
\newcommand{\mytag}{\stepcounter{mysubequation}%
\tag{\theequation\protect\sep{\themysubequation}}}
\newcommand{\globallabel}[1]{\refstepcounter{equation}\label{#1}}

\makeatletter
\renewcommand{\section}{\@startsection{section}{1}{0em}{-\baselineskip}%
{\baselineskip}{\normalfont\large\bfseries}}
\renewcommand{\subsection}%
{\@startsection{subsection}{2}{0em}{-0.7\baselineskip}%
{0.7\baselineskip}{\normalfont\bfseries}}
\makeatother

\newcommand{\GeV}{\,\mathrm{GeV}}

\newcommand{\eV}{\,\mathrm{eV}}

\newcommand{\VeV}[2][]{#1\langle #2 #1\rangle} %% \big\Big\bigg\Bigg  

\newcommand{\Nup}{{N_{\mu^+}}}
\newcommand{\Num}{{N_{\mu^-}}}
\newcommand{\Nupm}{{N_{\mu^\pm}}}
\newcommand{\nuu}{n_{\mu^-}(\mu^-)}
\newcommand{\nuub}{n_{\mu^+}(\mu^+)}
\newcommand{\neu}{n_{\mu^+}(\mu^-)}
\newcommand{\neub}{n_{\mu^-}(\mu^+)}
\newcommand{\neuvac}{n^{\text{vac}}_{\mu^+}(\mu^-)}
\newcommand{\neubvac}{n^{\text{vac}}_{\mu^-}(\mu^+)}
\newcommand{\eres}{{E_{\text{res}}}}

\newcommand{\reu}{{\nu_e\rightarrow\nu_\mu}}
\newcommand{\reub}{{\bar{\nu}_e\rightarrow\bar{\nu}_\mu}}
\newcommand{\ruu}{{\nu_\mu\rightarrow\nu_\mu}}
\newcommand{\ruub}{{\bar{\nu}_\mu\rightarrow\bar{\nu}_\mu}}

\newcommand{\emin}{{E_{\text{min}}}}
\newcommand{\NKT}{{N_{\text{kT}}}}
\newcommand{\dm}[1]{{\Delta m^2_{#1}}}

\def\ltap{\ \raisebox{-.4ex}{\rlap{$\sim$}} \raisebox{.4ex}{$<$}\ }
\def\gtap{\ \raisebox{-.4ex}{\rlap{$\sim$}} \raisebox{.4ex}{$>$}\ }

\begin{document}

%%%%%%%%%%%%%%%%%%%%%%%%%%%%%%%%%%%%%%%%%%%%%%%%%%%%%%%%%%%%%%%%%%%%%
%%%%                     Title-page                              %%%%
%%%%%%%%%%%%%%%%%%%%%%%%%%%%%%%%%%%%%%%%%%%%%%%%%%%%%%%%%%%%%%%%%%%%%

\begin{titlepage}

% the footnote symbols are only redefined for the title page !
\renewcommand{\thefootnote}{\alph{footnote}}

\ \vspace*{-3.cm}
\begin{flushright}
  {\hfill TUM--HEP--365/99}\\
  {OUTP--99--69P}\\
  {SISSA 151/99/EP}\\
  {\ }
\end{flushright}

\vspace*{0.5cm}

\renewcommand{\thefootnote}{\fnsymbol{footnote}}
\setcounter{footnote}{-1}

{\begin{center} {\LARGE\bf Testing Matter Effects in Very Long Baseline 
Neutrino Oscillation  Experiments$^*$\footnote{\hspace*{-1.6mm}$^*$Work 
        supported in part by
        "Sonderforschungsbereich 375 f\"ur Astro-Teilchenphysik" der Deutschen
        Forschungsgemeinschaft, by the TMR Network under the EEC
        Contract No.~ERBFMRX--CT960090 
        and by the Italian MURST under the program
        ``Fisica Teorica delle Interazioni Fondamentali''.}}
\end{center}}
\renewcommand{\thefootnote}{\alph{footnote}}

\vspace*{.8cm}
{\begin{center} {\large{\sc 
                M.~Freund\footnote[1]{\makebox[1.cm]{Email:}
                Martin.Freund@physik.tu-muenchen.de},~  
                M.~Lindner\footnote[2]{\makebox[1.cm]{Email:}
                Manfred.Lindner@physik.tu--muenchen.de},~
                S.~T.~Petcov\footnote[3]{\makebox[1.cm]{Email:}
                petcov@he.sissa.it}$^,$\footnote[4]{Also at: INRNE,
                Bulgarian Academy of Sciences, 1789 Sofia, Bulgaria.}
                \vspace{0.2cm}
                and
                {\sc 
                ~A.~Romanino\footnote[5]{\makebox[1.cm]{Email:}
                romanino@thphys.ox.ac.uk}}}}
\end{center}}
\vspace*{0cm}
{\it 
\begin{center}  

   \footnotemark[1]${}^,$\footnotemark[2]%
                Theoretische Physik, Physik Department, 
                Technische Universit\"at M\"unchen,\\
                James--Franck--Strasse, D--85748 Garching, Germany
   \vskip .3cm

   \footnotemark[3]%
                Scuola Internazionale Superiore di Studi Avanzati,
                and INFN - Sezione de Trieste, I-34014 Trieste, Italy
   \vskip .3cm

   \footnotemark[5]%
                Department of Physics, Theoretical Physics, University of
                Oxford,\\ Oxford OX13NP, UK

\end{center} }
\vspace*{0.5cm}
{\Large \bf 
\begin{center} Abstract \end{center}  }
Assuming three-neutrino mixing, we study the capabilities 
of very long baseline neutrino oscillation experiments to 
verify and test the MSW effect and to measure the lepton 
mixing angle $\theta_{13}$. We suppose that intense neutrino 
and antineutrino beams will become available in so--called 
neutrino factories. We find that the most promising and 
statistically significant results can be obtained by studying 
$\reu$ and $\reub$ oscillations which lead to matter 
enhancements and suppressions of wrong sign muon rates.
We show the $\theta_{13}$ ranges where matter effects could 
be observed as a function of the baseline. We discuss 
the scaling of rates, significances and sensitivities with 
the relevant mixing angles and experimental parameters. 
Our analysis includes fluxes, event rates and statistical 
aspects so that the conclusions should be useful for the 
planning of experimental setups. We discuss the subleading 
$\dm{21}$ effects in the case of the LMA~MSW solution of the 
solar problem, showing that they are small for 
$L\gtrsim 7000\,\text{km}$. For shorter baselines, $\dm{21}$ 
effects can be relevant and their dependence on $L$ offers 
a further handle for the determination of the CP-violation 
phase $\delta$. Finally we comment on the possibility to 
measure the specific distortion of the energy spectrum due 
to the MSW effect. 

\vspace*{.5cm}

\end{titlepage}

\newpage

\renewcommand{\thefootnote}{\arabic{footnote}}
\setcounter{footnote}{0}

%%%%%%%%%%%%%%%%%%%%%%%%%%%%%%%%%%%%%%%%%%%%%%%%%%%%%%%%%%%%%%%%%%%%%
%                     Introduction                                  %
%%%%%%%%%%%%%%%%%%%%%%%%%%%%%%%%%%%%%%%%%%%%%%%%%%%%%%%%%%%%%%%%%%%%%

\section{Introduction \label{sec:SEC-intro}}

The long term aim to build muon colliders offers the very 
attractive intermediate possibility for ``neutrino factories''
\cite{GEER98,gavela-c,BARGER99} with uniquely intense and precisely 
characterized neutrino and antineutrino beams. This requires only 
one muon beam at intermediate energies such that neutrino factories 
are rather realistic medium term projects which constitute also 
a useful step in accelerator technology towards a muon collider.
The current knowledge of neutrino masses and mixing implies 
for typical setups very promising very long baseline neutrino 
experiments. We study in this paper in a three neutrino framework 
the potential to verify and test the MSW effect and to measure or 
limit $\theta_{13}$ in terrestrial very long baseline experiments
with neutrino factories where the neutrino spectrum 
and fluxes are rather well known and under control \cite{GEER98}. 
Calculating the oscillation probabilities and event rates for 
different channels and comparing with those for oscillations in 
vacuum we find that the asymmetry between the 
$\nu_e\leftrightarrow\nu_\mu$ and 
$\bar\nu_e\leftrightarrow\bar\nu_\mu$ 
oscillations is a very promising tool to test and verify the MSW 
effect. The reason is, as we will see, that matter effects lead 
to measurably enhanced event rates in one of the two channels and to
equally suppressed transitions in the other. 
The asymmetry between the event rates associated
with these two channels would therefore be very sensitive to 
the MSW effect since the matter induced changes have opposite 
effect on the two rates thus amplifying the ``signal'', while 
at the same time common backgrounds would drop out. We analyze 
event rates and we include statistical aspects such that the 
results are directly applicable for the planning of optimal 
experimental setups. We discuss the capabilities of a 
neutrino-factory experiment as a function of the distance 
between the neutrino source and the detector and of the muon 
source energy for the optimal observation of the MSW effect. 
Moreover, we determine the sensitivity to the value
of $\theta_{13}$ for different experimental configurations.

Demonstrating and testing the MSW effect directly is of 
fundamental importance since this effect plays a basic role 
in different neutrino physics scenarios. The MSW mechanism 
provides, for example, the only clue for understanding the 
solar neutrino deficit with a neutrino mass squared difference 
within a few orders of magnitude from that inferred from 
the atmospheric neutrino data. Atmospheric neutrinos
can undergo matter-enhanced transitions in the earth.
The matter effects in neutrino oscillations will play an 
important role in the interpretation of the results of a neutrino 
factory experiment using a $L\gtrsim 1000\,\text{km}$ baseline. 
They also play a role in the searches for CP-violation in such 
experiments ~\cite{gavela-c,DFLR}, since matter effects 
generate an asymmetry between the two relevant CP-conjugated 
appearance channels \cite{DFLR}. Knowing the asymmetry 
caused by matter effects is therefore essential for obtaining 
information on the CP-violation originating from the lepton 
mixing matrix. Matter enhanced neutrino transitions can 
play important role in astrophysics as well. 

Neutrino factories are extensively discussed in the
literature~\cite{GEER98,gavela-c,BARGER99,BCR9905240,
gavela-b,romanino,BARGER99b}. Either muons or anti-muons are 
accelerated to an energy $E_\mu$ and decay then in straight 
sections of a storage ring like $\mu^{-} \rightarrow e^{-}
+ \bar{\nu}_e + \nu_{\mu}$ or $\mu^{+} \rightarrow e^{+} +
\bar{\nu}_{\mu} + \nu_{e}$ so that a very pure neutrino beam
containing $\bar{\nu}_e$ and $\nu_{\mu}$ or $\nu_e$ and
$\bar{\nu}_{\mu}$, respectively, is produced.  The muon 
energy $E_\mu$ could be in a wide range from $10\GeV$ to $50\GeV$ 
or more and a neutrino flux corresponding to $2\cdot 10^{20}$ 
muon decays per year in the straight section of the ring 
pointing to a remote detector could be achieved. Higher 
fluxes are also currently under discussion~\cite{lyon}. The
neutrino fluxes are therefore very intense and can be easily
calculated from the decay spectrum at rest. For unpolarized 
muons and negligible beam divergence one finds for a baseline 
of $L=730$~km a neutrino flux of 
$\simeq 4.3\cdot10^{12}~{\rm yr}^{-1}{\rm m}^{-2}$ and 
for a baseline $L=10000$~km a neutrino flux 
$\simeq 2.3\cdot10^{10}~{\rm yr}^{-1}{\rm m}^{-2}$. Note also that the
$\bar\nu_e$ and $\nu_e$ fluxes depend sizably on the beam
polarization, that will be assumed to be negligible in this 
paper. Altogether a neutrino factory would provide pure and 
high-intensity neutrino beams with a well known energy spectrum 
that in turn would allow a wide physical program including 
precise measurement of mixing parameters~\cite{gavela-c,BARGER99}, 
matter effects~\cite{BCR9905240,BARGER99b} and, in case of 
LMA solution of the solar problem, leptonic 
CP-violation~\cite{gavela-b,romanino}. 

The produced neutrino beam will be directed towards a 
remote detector at a given Nadir angle $h$, which corresponds 
to an oscillation baseline $L = 2R_{\oplus}\cos(h)$~, where 
$R_{\oplus} = 6371~$km is the earth radius. 
If $L \gtap 10^{3}~$km, as we shall assume, matter effects 
become important in neutrino oscillations\footnote{The possibility 
to detect matter effects in long baseline neutrino oscillation 
experiments with $L \simeq 730$~km (MINOS, CERN - GS) was discussed, 
e.g., in refs.~\cite{Lipari,Mohan}.}. For $L \leq 10600~$km 
the beam traverses the earth along a trajectory in the earth mantle 
without crossing the earth core where the 
density is substantially higher. According to the earth models 
\cite{Stacey77,PREM81}, the average matter density along the neutrino 
trajectories with $L = (10^{3} - 10^{4})~$km lies in the interval
$\sim$ (2.9 - 4.8) g/cm$^3$. The matter density changes along each 
trajectory, but the variation is relatively small - by about 1.5 
to 2.0 g/cm$^3$, and even the largest takes place over relatively 
big distances of several thousand kilometers. 
As a consequence, one can approximate the earth mantle density profile 
by a constant average density distribution. The constant density 
should be chosen to be equal to the average density along every 
trajectory \cite{SP1,MLiuSP96,MartinTommy}. For the calculation of 
the neutrino oscillation probabilities in the case of interest, the 
indicated constant density model provides a very good approximation 
to the somewhat more complicated density structure of the earth mantle 
(for a recent discussion see, e.g., \cite{Shrock99}).
Let us note also that the earth mantle is with a good precision
isotopically symmetric: $Y_e = 0.494$ \cite{Stacey77,PREM81},
where $Y_e$ is the electron fraction number in the mantle. 

Since the beam consists always either of $\nu_e$ together 
with $\bar\nu_\mu$ or $\bar \nu_e$ in combination with 
$\nu_\mu$ there are, in principle, eight different appearance 
experiments and four different disappearance experiments 
which could be performed\footnote{For pre-neutrino factory 
discussions of very long baseline terrestrial neutrino oscillation 
experiments see, e.g., \cite{MannP77}.}. From an experimental 
point of view, however, at present the four channels with muon neutrinos 
or antineutrinos in the final state, $\ruu$, $\ruub$, 
$\reu$, $\reub$ seem to be most promising. A very important 
issue is in this context the ability of the detector to 
discriminate between neutrinos and antineutrinos, 
namely the ability to measure the charge of the leptons 
produced by the neutrino charged current interactions. 
Note that very good discrimination capability is required 
for a measurement of the appearance probabilities since 
they produce  ``wrong sign'' muon events in the detector
and have to be discriminated from the much larger number 
of events associated with the $\nu_{\mu}$ and $\bar{\nu}_{\mu}$ 
survival probabilities. The channels with 
electron neutrinos or antineutrinos in the final state 
are problematic from this point of view due to the
difficulty of telling $e^+$ from $e^-$ in a large 
high-density detector. On the contrary, a very good 
$\mu^+/\mu^-$ discrimination could be obtained in a large 
properly oriented magnetized detector \cite{magdetector}. 

The paper is organized as follows. 
In section~2 we give the analytic formulae for three neutrino 
oscillations in matter which contain the essential physics 
relevant for our study. In the following section~3 we discuss 
event rates, their parameter dependence, their scaling behaviour
and we show results from our numerical calculations.
In section~4 we define the sensitivity to matter effects in 
a statistical sense and discuss the results of our numerical
calculations including parameter uncertainties. 
This is followed by a discussion in section~5 of the effects 
of a non-vanishing $\dm{21}$. In section~6 the possibility 
of detecting matter effects by looking for the characteristic 
enhancement, the broadening and the shift of the dominating 
maximum of the event rate energy spectrum is discussed and 
we conclude in section~7. 

%%%%%%%%%%%%%%%%%%%%%%%%%%%%%%%%%%%%%%%%%%%%%%%%%%%%%%%%%%%%%%%%%%%%%
%%%%       SECTION: three nu oscillation probabilities           %%%%
%%%%%%%%%%%%%%%%%%%%%%%%%%%%%%%%%%%%%%%%%%%%%%%%%%%%%%%%%%%%%%%%%%%%%

\section{Three Neutrino Oscillation Probabilities in Matter 
\label{sec:SEC-3nuinmatter}} 

We will assume in this paper the existence of three flavour 
neutrino mixing:
\begin{equation}
|\nu_l>~ = \sum_{k=1}^3 U_{lk}|\nu_k>~, 
\hspace{1cm} l=e,\mu,\tau, 
\label{numixing}
\end{equation}
where $|\nu_l>$ is the state vector of the (left-handed) flavour 
neutrino $\nu_l$, $|\nu_k>$  is the state vector of a neutrino
$\nu_k$ possessing a definite mass $m_k$, $m_k \neq m_j$,
$k \neq j = 1,2,3$ and $U$ is a $3 \times 3$ unitary matrix -- the lepton 
mixing matrix. We conventionally order the masses in such 
a way that $0 < \dm{21} < |\dm{31}|$. According to the sign of 
$\dm{31}$, the previous convention corresponds to the two closest neutrino 
masses ($m_1$,$m_2$) being lighter than the third one ($m_3$) 
($m_3>m_2> m_1$ and $\dm{31}>0$) or being heavier than the third one 
($m_2\gtrsim m_1 > m_3$ and $\dm{31}<0$)\footnote{This scheme implies 
an approximate degeneracy of either $m_1$ and $m_2$ or of all the mass 
eigenstates.}. As long as matter and CP-violation effects are not taken 
into account, the two cases are phenomenologically equivalent from the 
point of view of neutrino oscillations. It is natural to suppose 
in this case that one of the two independent neutrino mass-squared 
differences, say $\Delta m^2_{21}$, is relevant for the vacuum 
oscillation (VO), small or large mixing angle MSW solutions\footnote{There 
exists also a large mixing angle solution of the solar neutrino problem
with $\Delta m^2_{21} \sim 10^{-7}\eV^2~$ - the LOW solution, which, however,
provides a somewhat worse description of the data than the MSW and the VO 
solutions \cite{solardm2}.} (SMA~MSW and LMA~MSW) of the solar neutrino problem 
with values in the intervals \cite{solardm2,atmdm2}
\globallabel{Dm2}
\begin{align}
{\rm VO} & :~~~~5.0\times10^{-11}\eV^2\ltap \Delta m_{21}^2 \ltap
5.0\times 10^{-10}\eV^2~, \mytag \\ 
{\rm SMA~MSW} & :~~~~4.0\times10^{-6}\eV^2\ltap \Delta m_{21}^2 \ltap
9.0\times 10^{-6}\eV^2~,\mytag \\ 
{\rm LMA~MSW} & :~~~~2.0\times10^{-5}\eV^2\ltap \Delta m_{21}^2 \ltap
2.0\times 10^{-4}\eV^2~,\mytag
\end{align}
while $\Delta m^2_{31}$ is responsible for the dominant 
atmospheric $\nu_{\mu} \leftrightarrow \nu_{\tau}$ oscillations 
and lies in the interval  
\begin{equation}
{\rm ATM} :~~~~10^{-3}\eV^2\ltap |\Delta m_{31}^2| \ltap
8.0\times 10^{-2}\eV^2~.
\label{Dm2b}
\end{equation}
For  $E_{\nu} \geq 1\GeV$, $L \leq 10^{4}~$km, the $\Delta m^2$-hierarchy 
\begin{equation} 
\Delta m^2_{21} \ll |\Delta m^2_{31}|
\label{hierarchy}
\end{equation}
and $\Delta m^2_{21} \ll 10^{-4}\eV^2$, the probabilities of 
three-neutrino oscillations in vacuum of interest reduce effectively 
to two-neutrino vacuum oscillation probabilities \cite{ADeR80,BPet87}:
\begin{equation}
P^{3\nu}_{vac}(\nu_l \rightarrow \nu_{l'}) = 
P^{vac}(\bar{\nu}_l \rightarrow \bar{\nu}_{l'}) \cong
2|U_{l3}|^2 |U_{l'3}|^2~\left ( 1 -
\cos {\Delta m^2_{31}L \over {2E}} \right )~,~l \neq l'=e,\mu,\tau,  
\label{Pmu2tau}
\end{equation}
\begin{equation}
P^{3\nu}_{vac}(\nu_{l} \rightarrow \nu_{l}) = 
P^{3\nu}_{vac}(\bar{\nu}_{l} \rightarrow \bar{\nu}_{l}) \cong
1 - 2|U_{l3}|^2(1 - |U_{l3}|^2)~\left ( 1 -
              \cos {\Delta m^2_{31}L \over {2E}} \right ),~l=e,\mu,\tau.
\label{Pmue2emu}
\end{equation}
Under the conditions (\ref{numixing}) and (\ref{hierarchy}) the 
element $|U_{e3}|$ of the lepton mixing matrix, which controls the 
$\nu_e \rightarrow \nu_{\mu (\tau)}$,
$\bar{\nu}_e \rightarrow \bar{\nu}_{\mu (\tau)}$,
$\nu_{\mu} \rightarrow \nu_{e}$ and the
$\bar{\nu}_{\mu} \rightarrow \bar{\nu}_{e}$
oscillations, is tightly constrained by the CHOOZ experiment 
\cite{CHOOZ99} and the oscillation interpretation of the solar 
and atmospheric neutrino data: for $3.0\times 10^{-3}\eV^2 
\leq  \Delta m^2_{31} \leq 8.0\times 10^{-3}\eV^2$ one has
\begin{equation}
|U_{e3}|^2 \ltap 0.025~.
\label{Ue3}
\end{equation}
The CHOOZ upper limit is less stringent for  
$1.0\times 10^{-3}\eV^2 
\leq  |\Delta m^2_{31}| < 3.0\times 10^{-3}\eV^2$ where 
values of $~|U_{e3}|^2 \cong 0.05$ are allowed.

Note that under the condition (\ref{hierarchy}),
the VO or MSW transitions of solar neutrinos depend on 
$~|U_{e3}|^2$ and on the two-neutrino VO or MSW transition 
probability with $\Delta m^2_{21}$ and $\theta_{12}$, where 
\begin{equation}
\sin^22\theta_{12} = 4 {{|U_{e1}|^2~|U_{e2}|^2}\over {(|U_{e1}|^2 +
|U_{e2}|^2)^2}},~~
\cos 2\theta_{12} =  {{|U_{e1}|^2 - |U_{e2}|^2}\over {|U_{e1}|^2 +
|U_{e2}|^2}}~,
\label{trigonometry}
\end{equation}
playing the role of the corresponding two-neutrino oscillation 
parameters \cite{CSL87,3nuSP88,3nu} (for recent reviews see, e.g., 
\cite{SPWIN99,OY99}). For  $|U_{e3}|^2$ satisfying the limit 
(\ref{Ue3}), however, this dependence is rather weak and cannot 
be used to further constrain or determine $|U_{e3}|^2$.  
In general, under the condition (\ref{hierarchy}) and for 
$\Delta m^2_{21} \ll 10^{-4}\eV^2$ the relevant solar 
neutrino transition probability depends only on the absolute 
values of the elements of the first row of the lepton mixing 
matrix, i.e., on $|U_{ei}|^2$, i=1,2,3, while the 
oscillations of the (atmospheric) $\nu_{\mu}$,
$\bar{\nu}_{\mu}$, $\nu_e$ and $\bar{\nu}_e$ on earth distances 
are controlled by the elements of the third column
of $U$, $|U_{l3}|^2$, $l=e,\mu,\tau$. The other elements 
of $U$ are not accessible to direct experimental determination.
Moreover, the $CP-$ and $T-$ violation effects in the oscillations 
of neutrinos are negligible. 

For our analysis we use a standard 
parametrization of the lepton mixing matrix $U$:
\begin{equation}
\left(\begin{array}{ccc}
U_{e1}& U_{e2} & U_{e3} \\
U_{\mu 1} & U_{\mu 2} & U_{\mu 3} \\
U_{\tau 1} & U_{\tau 2} & U_{\tau 3} 
\end{array} \right)
= \left(\begin{array}{ccc} 
c_{12}c_{13} & s_{12}c_{13} & s_{13}e^{-i\delta}\\
- s_{12}c_{23} - c_{12}s_{23}s_{13}e^{i\delta} & 
 c_{12}c_{23} - s_{12}s_{23}s_{13}e^{i\delta} & s_{23}c_{13}\\
s_{12}s_{23} - c_{12}c_{23}s_{13}e^{i\delta} 
& -c_{12}s_{23} - s_{12}c_{23}s_{13}e^{i\delta}
& c_{23}c_{13}\\ 
\end{array} \right)
\label{Umix}
\end{equation}
where $c_{ij} \equiv \cos \theta_{ij}$, $s_{ij} \equiv \sin \theta_{ij}$ 
and $\delta$ is the Dirac CP-violation phase\footnote{We have not 
written explicitly the possible Majorana CP-violation phases 
which do not enter into the expressions for the oscillation 
probabilities of interest \cite{BHP80}. Furthermore we assume 
throughout this study $0 \leq \theta_{12}, \theta_{23}, 
\theta_{13} < \pi/2,~ 0 \leq \delta < 2\pi$.}. The angles $\theta_{12}$ 
and $\theta_{23}$ in (\ref{Umix}) are constrained to lie within rather
narrow intervals by the solar and  atmospheric neutrino data for 
each of the different solutions, eqs.~(\ref{Dm2}a) - (\ref{Dm2}c), 
of the solar neutrino problem. With the accumulation of data the 
uncertainties in the knowledge of  $\theta_{12}$ and $\theta_{23}$ 
will diminish, while only upper limits on $s^2_{13}$ like 
eq.~(\ref{Ue3}) have been obtained so far. The mixing angle 
$\theta_{13}$ is one of the 4 (or 6 - depending on whether the massive 
neutrinos are of Dirac or Majorana type \cite{BHP80,BPet87})
fundamental parameters in the lepton mixing matrix. It controls the 
probabilities of the $\nu_{\mu} (\nu_e) \rightarrow \nu_{e} (\nu_{\mu})$,
$\bar{\nu}_{\mu} (\bar{\nu}_e) \rightarrow \bar{\nu}_{e} (\bar{\nu}_{\mu})$,
$\nu_e \rightarrow \nu_{\tau}$ and $\bar{\nu}_e\rightarrow\bar{\nu}_{\tau}$ 
oscillations and the $CP-$ and $T-$violation effects in neutrino 
oscillations depends on it. Obviously, one of the main 
goals of the future neutrino oscillation experiments should be to 
determine the value of $\theta_{13}$ or to obtain a more stringent 
experimental upper limit than the existing one (\ref{Umix}).

Under the condition (\ref{hierarchy}), with 
$\Delta m^2_{21} \ll 10^{-4}\eV^2$ and in constant density 
approximation, the oscillation probabilities of interest take the 
following simple form (see, e.g., \cite{3nuSP88,s5398,OY99}):
\globallabel{Ps}
\begin{align}
P^{3\nu}_{E}(\nu_{\mu} \rightarrow \nu_{e})
 = & ~ P^{3\nu}_{E}(\nu_e \rightarrow \nu_{\mu}) \cong s_{23}^2~
P^{2\nu}_{E}(\Delta m^2_{31}, \sin^22\theta_{13})~, \mytag \\
P^{3\nu}_{E}(\nu_{\mu} \rightarrow \nu_{\mu}) \cong &
~ c^4_{23} + s^4_{23}\left [ 1 - P^{2\nu}_{E}(\Delta m^2_{31},
\sin^22\theta_{13}) \right ] \nonumber \\
& + 2c^2_{23}s^2_{23}~ \left [ \cos \kappa - (1 + \cos 2\theta^m_{13})~
\sin{{\Delta E_{m} L}\over 2}~ 
\sin(\kappa + {{\Delta E_{m} L}\over 2}) \right ]~,\mytag\\
P^{3\nu}_{E}(\nu_{\mu} \rightarrow \nu_{\tau})  \cong &
~ 2c^2_{23} s^2_{23}~\left [ 2\sin^2 {\kappa\over 2}
- {1\over 2}P^{2\nu}_{E}(\Delta m^2_{31},
\sin^22\theta_{13})\right. \nonumber \\
& + \left.  (1 + \cos 2\theta^m_{13})~ \sin{{\Delta E_{m} L}\over 2}~ 
\sin(\kappa + {{\Delta E_{m} L}\over 2}) \right ]~,\mytag
\end{align}
where
\begin{equation}
P^{2\nu}_{E}(\Delta m^2_{31}, \sin^22\theta_{13}) =  {1\over {2}} 
\left [1 - \cos \Delta E_{m} L \right ] \sin^2 2\theta_{13}^{m}
\label{PE2nu}
\end{equation} 
is the two-neutrino transition probability in matter with constant 
density and
\begin{equation}
\kappa \cong {L\over {2}}~\left[ {\Delta m^2_{31}\over{2E}} 
+ V - \Delta E_{m} \right]
\label{kappa}
\end{equation}
is a phase. In eqs.~(\ref{PE2nu}) - (\ref{kappa}), $\Delta E_{m}$ 
and $\theta^m_{13}$ are the neutrino energy difference and mixing 
angle in matter,
\begin{equation}
\Delta E_{m} = {|\Delta m^2_{31}|\over{2E}}C_{+}, \quad 
\cos 2\theta^m_{13} = {1\over{C_{+}}}\left (\cos2\theta_{13} - 
{{2EV}\over{\Delta m^2_{31}}} \right )~,
\label{Emcos2thetam}
\end{equation}
where 
\begin{equation}
C^2_{\pm} = \left(1 \mp \frac{2EV}{\Delta m^2_{31}}\right)^2 
\pm 4\frac{2EV}{\Delta m^2_{31}}\sin^2\theta_{13}
\label{cpm}
\end{equation}
and 
\begin{equation}
V = \sqrt{2} G_F \bar{N}_e^{man}
\label{matt}
\end{equation}
is the matter term, $\bar{N}_e^{man}$ being the average electron 
number density along the neutrino trajectory in the earth mantle,
The probability $P^{3\nu}_{E}(\nu_e \rightarrow \nu_{\tau})$
can be obtained from eq.~(\ref{Ps}a) by replacing the factor 
$s_{23}^2$ with $c_{23}^2$. The corresponding antineutrino
transition and survival probabilities have the same form and 
can formally be obtained from eqs.~(\ref{Ps}a--\ref{Ps}c) 
by changing the sign of the matter term, i.e.\
$\bar{N}_e^{man} \rightarrow - \bar{N}_e^{man}$
in the expressions for $\kappa$, $\Delta E_{m}$ 
$\cos 2\theta^m_{13}$, eqs.~(\ref{kappa})--(\ref{Emcos2thetam}),
and by replacing $C_{+}$ by $C_{-}$. Expressions (\ref{Ps}) follow 
directly from eqs.~(12), (13b), (14b) and eq.~(19) in 
\cite{3nuSP88} (after setting in the corresponding 
formulae $\varphi'_{23} = 0$, $\varphi'_{12} = \pi/2$). They can 
be deduced also from the explicit expressions for the probabilities 
of neutrino oscillations of the earth-core-crossing neutrinos
in the two-layer approximation for the earth density distribution 
(see, e.g., \cite{SP1,MLiuSP96,MartinTommy,KP3nu88}), obtained in 
\cite{s5398} (on the basis of \cite{3nuSP88}) and in \cite{ADLS99}. 

As we shall see in Section 5, at distances $L \lesssim 6000$ km 
and for $\Delta m^2_{21} \sim 10^{-4}~{\rm eV^2}$, the 
$\Delta m^2_{21}$ corrections in the probabilities of interest 
$P^{3\nu}_{E}(\nu_{\mu} \rightarrow \nu_{e})$,
$P^{3\nu}_{E}(\nu_{\mu} \rightarrow \nu_{\mu})$ and
$P^{3\nu}_{E}(\nu_{\mu} \rightarrow \nu_{\tau})$ can be non-negligible.
Utilizing the results of \cite{3nuSP88} it is not difficult to
derive also the expressions for these probabilities including the 
leading order (CP-conserving and CP-violating) 
$\Delta m^2_{21}$-corrections \cite{SPtbp}. One finds in the case 
of the probability $P^{3\nu}_{E}(\nu_{e} \rightarrow \nu_{\mu})$:
\begin{equation}
\begin{align}
P^{3\nu}_{E}(\nu_{e} \rightarrow \nu_{\mu})
& \cong s_{23}^2\left [ 1 + 
 \cos\delta \cot \theta_{23} \sin2\theta'_{23}\right ]
P^{2\nu}_{E}(~\overline{\Delta m}^2_{31}, \theta_{13})~ \nonumber \\
&~+~ \cos\theta'_{12}~\sin2\theta_{23}~\sin 2\bar{\theta}_{13}^{m}
~\sin {{\Delta \bar{E}_{m} L}\over 2}~\nonumber \\ 
&~\times~ \left [ \cos\delta \left ( \sin (\bar{\kappa}  
+ {{\Delta \bar{E}_{m} L}\over 2}) - \cos 2\bar{\theta}_{13}^{m} 
\sin{{\Delta \bar{E}_{m} L}\over 2} \right )~ \right. \nonumber \\
& - \left. 2 \sin\delta~\sin {\bar{\kappa}\over 2}~
\sin ({\bar{\kappa}\over 2} + {{\Delta \bar{E}_{m} L}\over 2}) \right ]~, 
\label{Pemudelta}
\end{align}
\end{equation}
where \cite{3nuSP88,SPtbp}
\begin{equation}
\overline{\Delta m}^2_{31} \equiv \Delta m^2_{31} - s_{12}^2 \Delta m^2_{21}~,
\label{Delta31}
\end{equation}
\begin{equation}
\bar{\kappa} = {L\over {2}}~\left[ {~\overline{\Delta m}^2_{31}\over{2E}} 
+ V - \Delta \bar{E}_{m} \right] 
- { {L\Delta m^2_{21}}\over {2E}}\cos 2\theta_{12}
\label{bkappa}
\end{equation}
and 
\begin{equation}
\cos\theta'_{12} = {{\Delta m^2_{21} c_{12}s_{12} c_{13}}\over
{2EV + s^2_{13}~\overline{\Delta m}^2_{31} - 
\Delta m^2_{21}\cos2\theta_{12}}} \cong
{{\Delta m^2_{21} c_{12}s_{12} c_{13}}\over
{2EV + s^2_{13}~\Delta m^2_{31}}}~,
\label{12prime}
\end{equation}
\begin{equation}
\sin 2\theta'_{23} = - ~{{\Delta m^2_{21} s_{13}\sin2\theta_{12}}\over
{\overline{\Delta m}^2_{31}c_{13}^2 - \Delta m^2_{21}\cos2\theta_{12}}} 
\cong - ~{{\Delta m^2_{21}\over {\Delta m^2_{31}}}~s_{13}\sin2\theta_{12}}~.
\label{23prime}
\end{equation}
The expressions for $P^{2\nu}_{E}(~\overline{\Delta m}^2_{31}, \theta_{13})$,
$\Delta \bar{E}_{m}$ and $\cos 2\bar{\theta}_{13}^{m}$ are given by
eqs.~(\ref{PE2nu}) and (\ref{Emcos2thetam}) in which $\Delta m^2_{31}$
is replaced by $\overline{\Delta m}^2_{31}$. The probability
$P^{3\nu}_{E}(\bar{\nu}_{e} \rightarrow \bar{\nu}_{\mu})$
($P^{3\nu}_{E}(\nu_{\mu} \rightarrow \nu_{e})$) can be obtain from 
eq.~(\ref{Pemudelta}) by making the change $V \rightarrow -~V$ and 
$\delta \rightarrow -~\delta$ ($\delta \rightarrow -~\delta$).
The expression (\ref{Pemudelta}) was obtained by neglecting the 
$\sim (\cos \theta'_{12})^2,~(\sin 2\theta'_{23})^2,~
\cos \theta'_{12}\sin 2\theta'_{23}$ and the higher order 
corrections (for further details see \cite{3nuSP88,SPtbp}).
Note that if $\Delta m^2_{21} \leq 2\times 10^{-4}~{\rm eV^2}$ 
and for $\bar{N}_e^{man} \gtrsim 1.45~{\rm cm^{-3}N_{A}}$
($L \geq 1000$ km), we have $2EV \geq 10^{-3}~{eV^2}$, and 
consequently $|\cos\theta'_{12}| \leq 0.1$ at $E \gtrsim 4.5~$GeV.
For the indicated maximal value of $\Delta m^2_{21}$ and 
$|\Delta m^2_{31}| \geq 10^{-3}~{\rm eV^2}$,
$s^2_{13} \lesssim 0.025~(0.05)$ one finds
$|\sin 2\theta'_{23}| \lesssim 0.032~(0.045)$.
It follows from the preceding discussion that if 
$|\Delta m^2_{31}| \cong 3.5\times 10^{-3}~{\rm eV^2}$,
the correction $\sim \cos \theta'_{12}$ would be
the dominant one for $E \sim (5 - 30)$ GeV.
For the values of $\Delta m^2_{31}$ and $\Delta m^2_{21}$
of interest, the $\Delta m^2_{21}-$correction included in 
$\overline{\Delta m}^2_{31}$, eq.~(\ref{Delta31}),
essentially just shifts the resonance energy by at most $\sim$ 10\%.
The terms with the factors $\cos\delta$ and $\sin\delta$ in 
eq.~(\ref{Pemudelta}) include the leading order CP-conserving 
and CP-violating  $\Delta m^2_{21}-$corrections, associated with 
the phase $\delta$.

Several comments are in order. First, the analytic expressions 
eqs.~(\ref{Ps}a)--(\ref{Ps}c) represent excellent 
approximations in the case of the VO and SMA~MSW solutions of 
the solar neutrino problem and for values of 
$\Delta m^2_{21} \ltap 5\times 10^{-5}\eV^2$ from the
LMA~MSW solution region. For $5\times 10^{-5}\eV^2 < 
\Delta m^2_{21} \ltap 2\times 10^{-4}\eV^2$,
the corrections due to $\Delta m^2_{21}$ can be non-negligible
and we are going to discuss them in section~\ref{sec:sun}.
Second, as it is clear from eqs.~(\ref{Ps}a) and (\ref{PE2nu}), 
the probabilities $P^{3\nu}_{E}(\nu_e \rightarrow \nu_{\mu})$ 
and $P^{3\nu}_{E}(\bar{\nu}_e \rightarrow \bar{\nu}_{\mu})$
cannot exceed $s^2_{23} \sim 0.5$. The maximal value 
$s^2_{23}$ can be reached only if the MSW  resonance 
condition $\sin^22\theta^{m}_{13} = 1$ and the condition
$\cos \Delta E_{m} L = - 1$ are simultaneously fulfilled.
At the MSW resonance, however, one has
$\Delta E^{res}_{m} \cong 1.23 \pi \times 10^{-4}~{\rm km}^{-1}~
\tan 2\theta_{13}\bar{N}_e^{man}[N_{A}{\rm cm^{-3}}]$,
where $\bar{N}_e^{man}$ is in units of 
$N_{A}{\rm cm^{-3}}$, $N_{A}$ being the Avogadro number,
and for $s^2_{13} \leq 0.025$ the second condition
$\cos \Delta E^{res}_{m}L = -1$ can only be satisfied for 
$L \geq 10^{4}~$km. If $s^2_{13} = 0.05$ this condition 
requires $L \geq 8\times 10^{3}~$km. Given the fact that the 
neutrino fluxes decrease with the distance as $L^{-2}$, the 
above discussion suggests that the maximum of the event 
distribution in $E$ and $L$ in the transition 
$\nu_e \rightarrow \nu_{\mu}$ should take place  approximately
at the MSW resonance energy, but at values of $L$ smaller 
than those at which $~max\left(P^{3\nu}_{E}(\nu_e \rightarrow 
\nu_{\mu})\right) \cong s^2_{23}$.  The MSW resonance energy is 
given by: $\eres \cong 6.56 \times \Delta m^2_{31}[10^{-3}\eV^2] 
\cos 2\theta_{13} (\bar{N}_e^{man}[cm^{-3}N_{A}])^{-1}\GeV$, 
where $\Delta m^2_{31}$ is in units of $10^{-3}\eV^2$. 
For $\Delta m^2_{31}[10^{-3}\eV^2] = 3.5;6.0,8.0$ and, e.g.,
$\bar{N}_e^{man}[cm^{-3}N_{A}] = 2$, we have 
$\eres \cong 11.5;19.7;26.2 \cos 2\theta_{13}\GeV$. The 
$\nu_e\rightarrow\nu_\mu$ and $\bar{\nu}_e\rightarrow\bar{\nu}_\mu$ 
oscillations will be affected substantially by the MSW effect if 
the energy of the parent muon beam $E_{\mu} > \eres$.
This condition can be satisfied for any value of 
$\Delta m^2_{31}$ from the interval~(\ref{Dm2b}) for 
$E_{\mu} \gtap 30\GeV$. For the value of 
$\Delta m^2_{31}[10^{-3}\eV^2] = 3.5$ which is currently 
``preferred'' by the Super-Kamiokande data we have 
$E_{\mu} \gtap 15\GeV$. Note that a change in the sign of
$\Delta m^2_{31}$ reverses the sign of $N^{res}_e$ which opens the
possibility to determine this sign via matter effects, as was noticed
recently also in ref.~\cite{BARGER99b}. The probabilities (\ref{Ps}) 
for $L = 7330$ km (the Fermilab - Gran Sasso distance) have been 
discussed recently in \cite{Shrock99}.

%%%%%%%%%%%%%%%%%%%%%%%%%%%%%%%%%%%%%%%%%%%%%%%%%%%%%%%%%%%%%%%%%%%%%
%%%%                     SECTION: Event rates                    %%%%
%%%%%%%%%%%%%%%%%%%%%%%%%%%%%%%%%%%%%%%%%%%%%%%%%%%%%%%%%%%%%%%%%%%%%

\section{Event Rates\label{sec:SEC-rates}} 

We discuss now the effects of matter on the total rate of 
$\reu$, $\reub$, $\ruu$, $\ruub$ events. If, for example, 
$\mu^+$ are accumulated in the storage ring, the 
neutrino beam contains $\nu_e$ and $\bar\nu_\mu$. Since
the neutrino-antineutrino transitions have a negligible rate, 
the beam-induced $\mu^+$ events in the detector must be 
attributed in this case to the charged current interactions 
of unoscillated $\bar\nu_\mu$. The total number $\nuub$ 
of $\mu^+$ events measures therefore an averaged $\bar\nu_\mu$ 
survival probability. Wrong sign $\mu^-$ events must be 
attributed to $\nu_\mu$ generated by oscillations of the 
initial $\nu_e$, so that their total number $\neu$ measures 
the averaged $\reu$ oscillation probability. Analogously, 
for $\mu^-$ in the storage ring, the total numbers of $\mu^-$ 
and $\mu^+$ events $\nuu$, $\neub$ measure the averaged 
$\nu_\mu$ survival probability and $\reub$ oscillation
probability, respectively.

The total number of events in each channel are given by
\globallabel{nunum}
\begin{align}
\neu &= \Nup \NKT \frac{10^9
N_A}{m^2_\mu\pi}\frac{E^3_\mu}{L^2}\int^{E_\mu}_\emin
f_{\nu_e\nu_\mu}(E) {P^{3\nu}_E}\text{\!\scriptsize $(\reu)$}(E)\,(dE/E_\mu)
~,\mytag \\ 
\neub &= \Num \NKT \frac{10^9
N_A}{m^2_\mu\pi}\frac{E^3_\mu}{L^2}\int^{E_\mu}_\emin
f_{\bar\nu_e\bar\nu_\mu}(E) {P^{3\nu}_E}\text{\!\scriptsize
$(\reub)$}(E)\,(dE/E_\mu) ~,\mytag \\ 
\nuu &= \Nup \NKT \frac{10^9
N_A}{m^2_\mu\pi}\frac{E^3_\mu}{L^2}\int^{E_\mu}_\emin
f_{\nu_\mu\nu_\mu}(E) {P^{3\nu}_E}\text{\!\scriptsize $(\ruu)$}(E)\,(dE/E_\mu)
~,\mytag \\ 
\nuub &= \Num \NKT \frac{10^9
N_A}{m^2_\mu\pi}\frac{E^3_\mu}{L^2}\int^{E_\mu}_\emin
f_{\bar\nu_\mu\bar\nu_\mu}(E) {P^{3\nu}_E}\text{\!\scriptsize
$(\ruub)$}(E)\,(dE/E_\mu)~, \mytag
\end{align}
where ${P^{3\nu}_E}$ denotes the oscillation probability described 
in section~\ref{sec:SEC-3nuinmatter}, $\Nup$ ($\Num$) is the number 
of  ``useful'' $\mu^+$ ($\mu^-$) decays, namely the number of decays 
occurring in the
straight section of the storage ring pointing to the detector, $\NKT$
is the size of the detector in kilotons, $10^9 N_A$ is the number of
nucleons in a kiloton, $E_\mu$ is the energy of the muons in the ring
and $\emin=3\GeV$ is a lower cut on the neutrino energies that helps
a good detection efficiency. Since low energy events are suppressed by
the low initial flux and the low cross section (see below), the results 
do not depend significantly on the precise value of $\emin$ for 
$\emin < 5\GeV$. The functions $f$ averaging the probabilities are given by
\globallabel{fdef}
\begin{align}
f_{\nu_e\nu_\mu}(E) &=g_{\nu_e}(E/E_\mu)
(\sigma_{\nu_\mu}(E)/E_\mu)\epsilon_{\mu^-}(E)~, \mytag \\
f_{\bar\nu_e\bar\nu_\mu}(E) &=g_{\bar{\nu}_e}(E/E_\mu)
(\sigma_{\bar{\nu}_\mu}(E)/E_\mu)\epsilon_{\mu^+}(E)~, \mytag \\
f_{\nu_\mu\nu_\mu}(E) &=g_{\nu_\mu}(E/E_\mu)
(\sigma_{\nu_\mu}(E)/E_\mu)\epsilon_{\mu^-}(E)~, \mytag \\
f_{\bar\nu_\mu\bar\nu_\mu}(E) &=g_{\bar{\nu}_\mu}(E/E_\mu)
(\sigma_{\bar{\nu}_\mu}(E)/E_\mu)\epsilon_{\mu^+}(E) \mytag
\end{align}
and take into account the appropriately normalized initial spectrum 
of $\nu$-neutrinos produced in the decay of unpolarized muons,
$g_\nu(E/E_\mu)$, the charged current cross section per nucleon,
$\sigma_{\nu_\mu (\bar{\nu}_\mu)}(E)$, and the efficiency for the
detection of $\mu^-$ ($\mu^+$), $\epsilon_{\mu^- (\mu^+)}(E)$ (we
neglect here the finite resolution of the detector).  For the
numerical calculations we use 
\globallabel{numeric}
\begin{align}
g_{\nu_e}(x) &= g_{\bar{\nu}_e}(x) = 12x^2(1-x)~, &
g_{\nu_\mu}(x) &= g_{\bar{\nu}_\mu}(x) = 2x^2(3-2x)~, \mytag \\
\sigma_{\nu_\mu}(E) &= 
0.67\cdot 10^{-38}E\,\text{cm}^2/\text{GeV}~, &
\sigma_{\bar{\nu}_\mu}(E) &= 
0.34\cdot 10^{-38}E\,\text{cm}^2/\text{GeV} \mytag
\end{align}
and $\epsilon_{\mu^-}(E)=\epsilon_{\mu^+}(E)=\epsilon$ for $E>\emin$
so that $f_{\nu_e\nu_\mu}(E)/f_{\bar\nu_e\bar\nu_\mu}(E) =
f_{\nu_\mu\nu_\mu}(E)/f_{\bar\nu_\mu\bar\nu_\mu}(E) = 2$ independently
of the energy. 

The contribution of the background to the number of muons observed 
in the detector has been neglected in eqs.~(\ref{nunum}). That 
background includes muons from the decay of charm quarks produced 
by charged and neutral current neutrino interactions in the detector 
and from the decay of $\tau$ produced by $\nu_{\tau}$ interactions. 
Both these sources can be kept under control in different 
ways~\cite{CDG,BARGER99b}. Let us note also that in eqs.~(\ref{nunum}) 
the possible divergence of the muon beam in the straight section of 
the storage ring has been neglected. 

We can analyze now the dependence of $\neu$, $\neub$, $\nuu$, $\nuub$ 
on the mixing angles, the CP-phase $\delta$ and some of the experimental 
parameters. In the $\Delta m^2_{12} \ll 10^{-4}\eV^2$ approximation 
discussed above, the only relevant mixing angles are $\theta_{23}$ and 
$\theta_{13}$. In the leading order in a $\theta_{13}$ expansion the
dependence of the total event rates in matter (and in vacuum) on these
parameters is 
\globallabel{laws}
\begin{align}
\nuu,\nuub &\propto\sin^2 2\theta_{23}~, \mytag \\
\neu,\neub &\propto\sin^2 \theta_{23}\sin^2 2\theta_{13}~. \mytag 
\end{align}
Higher order corrections in $\theta_{13}$ are constrained by the CHOOZ
limit. Despite the resonant matter enhancement of the mixing due to 
$\theta_{13}$, such corrections become only sizable for $\theta_{13}$
close to its upper limit and very long baselines where they can reach 
about 20\% in the resonance channels. 

\begin{figure}
\begin{center}
\epsfig{file=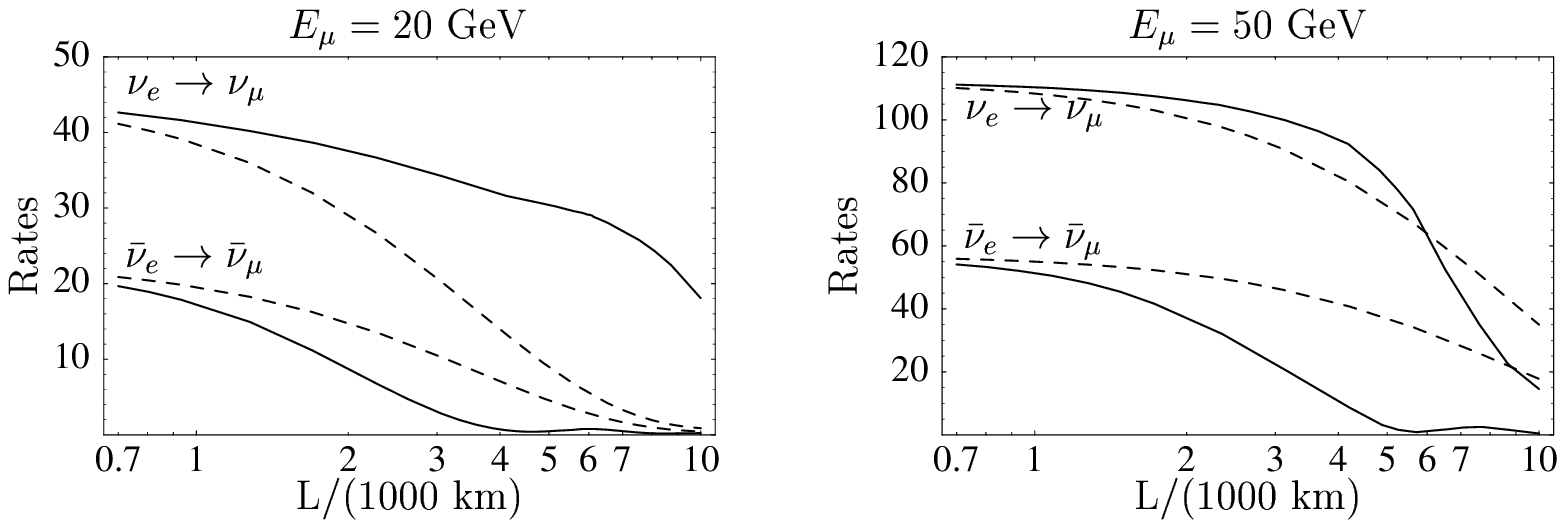,width=1.00\textwidth}
\end{center}
\mycaption{Appearance event rates $\neu$, $\neub$ due to the 
$\nu_e \rightarrow \nu_{\mu}$ and 
$\bar{\nu}_e \rightarrow \bar{\nu}_{\mu}$ transitions in the 
earth mantle (solid lines) and in vacuum (dashed lines), as 
functions of the baseline $L$ for $E_\mu=20\GeV$ (left) and 
$E_\mu=50\GeV$ (right). The differences between the solid and 
dashed lines is a measure of the magnitude of the earth matter 
effect. Both plots assume $N_\mu=2\cdot 10^{20}$, $\epsilon=50\%$,
$\sin^2 2\theta_{23}=1$ and $\sin^2 2\theta_{13}=0.01$. The scaling of
the rates with these parameters is described in the text.}
\label{fig:ratesap}
\end{figure}

\begin{figure}
\begin{center}
\epsfig{file=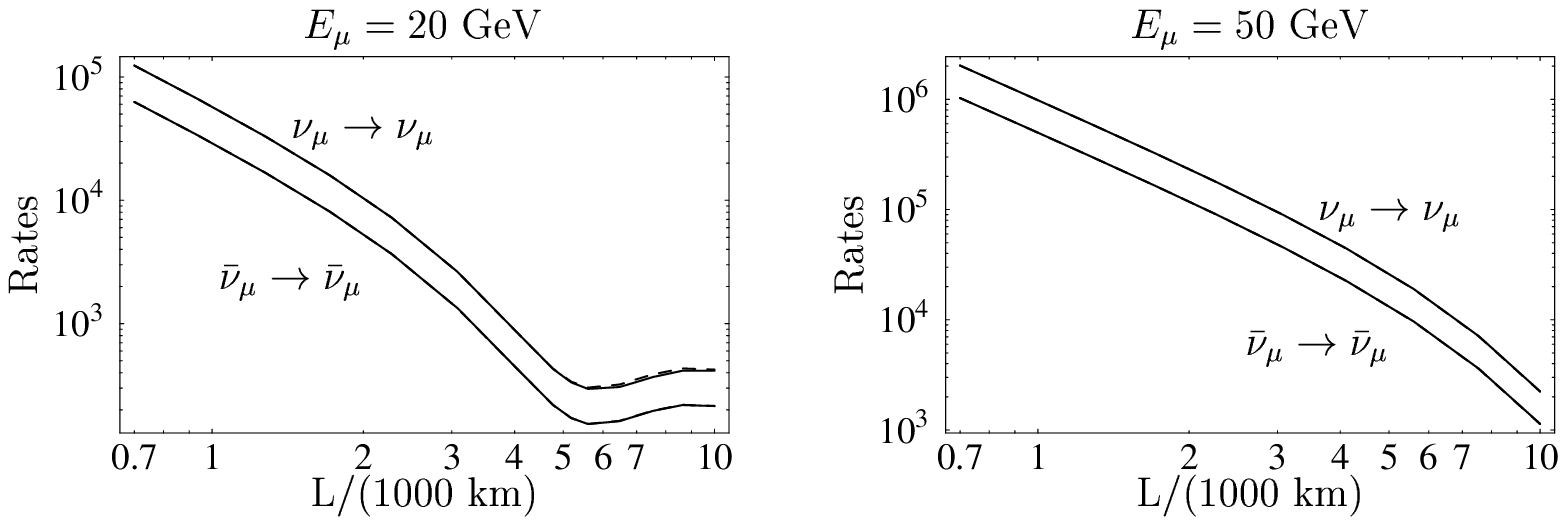,width=1.00\textwidth}
\end{center}
\mycaption{Same as in fig.~\ref{fig:ratesap} but for the disappearance
channels. The dashed lines coincide almost perfectly  with the solid 
lines, showing that matter effects are negligible in these channels.}
\label{fig:ratesdis}
\end{figure}

Since eq.~(\ref{laws}b) will turn out to be useful when discussing
the sensitivity to matter effects, we discuss it in greater detail. 
Let us write eqs.~(\ref{Ps}a) in the form
\begin{equation}
\label{pro}
P^{3\nu}_E(\reu,\reub) = \sin^2\theta_{23}\sin^22\theta_{13} 
\left(\frac{\sin(\Delta_{31} C_\pm)}{C_\pm}\right)^2,
\end{equation} 
where $C_+$ is given by eq.~(\ref{cpm}) and corresponds to
neutrinos, $C_-$ to antineutrinos and $\Delta_{31} = \dm{31}L/(4E)$.
Note that $C_+$ enhances the rates when $\dm{31} > 0$, depletes them
when $\dm{31} <0$. Therefore the sign of $\dm{31}$ determines whether
the enhanced channel is the neutrino or the antineutrino one.
In the limit in which $\sin^2\theta_{13}$ can be neglected 
on the right-hand side of eq.~(\ref{cpm}), eq.~(\ref{pro}) 
shows that $\neu$, $\neub$ are indeed proportional to 
$\sin^22\theta_{13}$. Despite the CHOOZ
limit, $\sin^2\theta_{13}\lesssim 0.025$, the second term in
eq.~(\ref{cpm}) can be relevant when the first term vanishes 
around the resonance. For $2EV = \Delta m^2_{31}$ we have in 
fact $C_+ = 2\sin\theta_{13}$ and
\begin{equation}
\label{pro1}
P^{3\nu}_E(\reu) =
\sin^2\theta_{23}\sin^22\theta_{13}\,\frac{L^2V^2}{4}
\left(\frac{\sin(\sin\theta_{13}LV)}{\sin\theta_{13}LV}\right)^2,
\end{equation}
whereas by neglecting the $\sin^2\theta_{13}$ term in~(\ref{cpm})
we would get $C_+ = 0$ and 
\begin{equation}
\label{pro2}
P^{3\nu}_E(\reu) =
\sin^2\theta_{23}\sin^22\theta_{13}\,\frac{L^2V^2}{4}~.
\end{equation}
A comparison of eqs.~(\ref{pro1}) and~(\ref{pro2}) shows however 
that this approximation works even at the resonance provided
$L\lesssim\pi/(4V\sin\theta_{13})\sim 7000\,
\text{km}\,(0.15/\sin\theta_{13})$. The effect of the
$\sin^2\theta_{13}$ term in eq.~(\ref{cpm}) is therefore maximal in
eq.~(\ref{cpm}) for very long baselines and $\theta_{13}$ close to the
experimental limit. In this case it can affect the rates sizably,
while it is negligible for smaller $\theta_{13}$ or smaller baseline.
For a better approximation one can use, for muon energies
such that the main contribution to the rates comes from neutrinos 
at the resonance, 
\begin{equation}
\label{lawb}
\neu,\neub \propto \sin^2\theta_{23}\sin^22\theta_{13}
\left(\frac{\sin(\sin\theta_{13}LV)}{\sin\theta_{13}LV}\right)^2~, 
\end{equation}
which deviates less than 5\% from the exact result in the whole
parameter space ($L\lesssim 10000\,\text{km}$, $20\GeV\lesssim E_\mu
\lesssim 50\GeV$). In most cases $L\lesssim\pi/(4V\sin\theta_{13})$
holds and the oscillating term in eq.~(\ref{lawb}) can be expanded,
giving eqs.~(\ref{laws}) or higher order approximations. We stress
that in eq.~(\ref{lawb}) the baseline $L$ appears as part of the
correction to the $\theta_{13}$, $\theta_{23}$ scaling only. The
dependence on $L$ of the rates is more involved, especially for very
large baselines, and will be described in figs.~\ref{fig:ratesap},
\ref{fig:ratesdis}.  Note that the dependence on the beam intensity,
detector size and efficiency is trivial: 
\globallabel{laws2}
\begin{align}
\neu,\nuu &\propto N_{\mu^{+}}\NKT\,\epsilon_{\mu^{+}}~, \mytag \\
\neub,\nuub &\propto N_{\mu^{-}}\NKT\,\epsilon_{\mu^{-}}~. \mytag
\end{align}

We present now quantitative results for the total rates in matter
for  $\Delta m^2_{21} \ll 10^{-4}\eV^2$ and we compare them with the 
results one would obtain in vacuum. The statistical significance 
of the matter effects will be discussed in the following section 
and effects of larger $\Delta m^2_{21}$ will be covered
in section~\ref{sec:sun}.

The total event rates depend, as already discussed, in a transparent 
way on the experimental parameters $\Nupm$, $\NKT$, $\epsilon_{\mu^\pm}$
and on the mixing parameters $\theta_{23}$, $\theta_{13}$. We
focus our discussion therefore on the less transparent dependence 
on the baseline and muon energy. For that we use the central value 
of $\Delta m^2_{31}=3.5\cdot 10^{-3}\eV^2$, and we assume
$\Delta m^2_{31}>0$. In the 
$\Delta m^2_{21} \ll 10^{-4}\eV^2$ approximation, in which
CP-violation effects are negligible, the results for $\Delta m^2_{31}<0$
can be obtained by simply interchanging neutrinos and antineutrinos.

The total number of events in the two appearance channels $\reu$,
$\reub$ is shown for $E_\mu=20\GeV$ and $E_\mu=50\GeV$ in
fig.~\ref{fig:ratesap} as a function of the baseline (solid
lines) in comparison with the event rates one would get if 
the neutrinos did not interact with matter (dashed lines).
Fig.~\ref{fig:ratesdis} shows the same results for the two 
disappearance channels $\ruu$, $\ruub$. Both figures correspond 
to a ``default'' experimental set-up providing 
$\Nup=\Num=N_\mu=2\cdot 10^{20}$ useful muon decays 
(e.g.\ in one year of running) and to a detector with $\NKT=10$ 
kilotons and an efficiency 
$\epsilon_{\mu^+}=\epsilon_{\mu^-}=\epsilon=50\%$ in both 
channels.  The rates depend only on the combination $N_\mu
\NKT\epsilon$ which is in our case $N_\mu \NKT\epsilon=10^{21}$. 
We assume for these figures that $\sin^22\theta_{23}=1$
and $\sin^2 2\theta_{13}=0.01$, one order of magnitude below 
the experimental limit. The rates for different values
of $N_\mu$, $\NKT$, $\epsilon$, $\theta_{23}$, $\theta_{13}$ can be
obtained by using eqs.~(\ref{laws2}), (\ref{laws}a) and~(\ref{laws}b)
or~(\ref{lawb}).


The vacuum event rates in the neutrino channels are twice as 
big as the rates in the antineutrino channels. This is because 
the oscillation probabilities are in the CP-conjugated channels  
in the $\Delta m^2_{21} \ll 10^{-4}\eV^2$ approximation the same 
while the functions averaging the probabilities are larger by a 
factor 2 in the neutrino channels (due to the larger cross-section). 
The disappearance channels shown in fig.~\ref{fig:ratesdis} are 
essentially independent of matter effects since these effects 
come only with the $\theta_{13}$ corrections to the oscillation 
probabilities\footnote{As discussed above effects as large as 
20\% can occur for $\theta_{13}$ close to the CHOOZ limit and 
very large baselines.}. 
In contrast, fig.~\ref{fig:ratesap} shows the drastic
enhancement (depletion) of the event rates in the $\reu$ ($\reub$)
channel for very long baselines. The growth of the total rates 
with the muon energy which is obvious from eqs.~(\ref{nunum}) 
can also be seen in fig.~\ref{fig:ratesap}. 

\begin{figure}[hb!]
\begin{center}
\epsfig{file=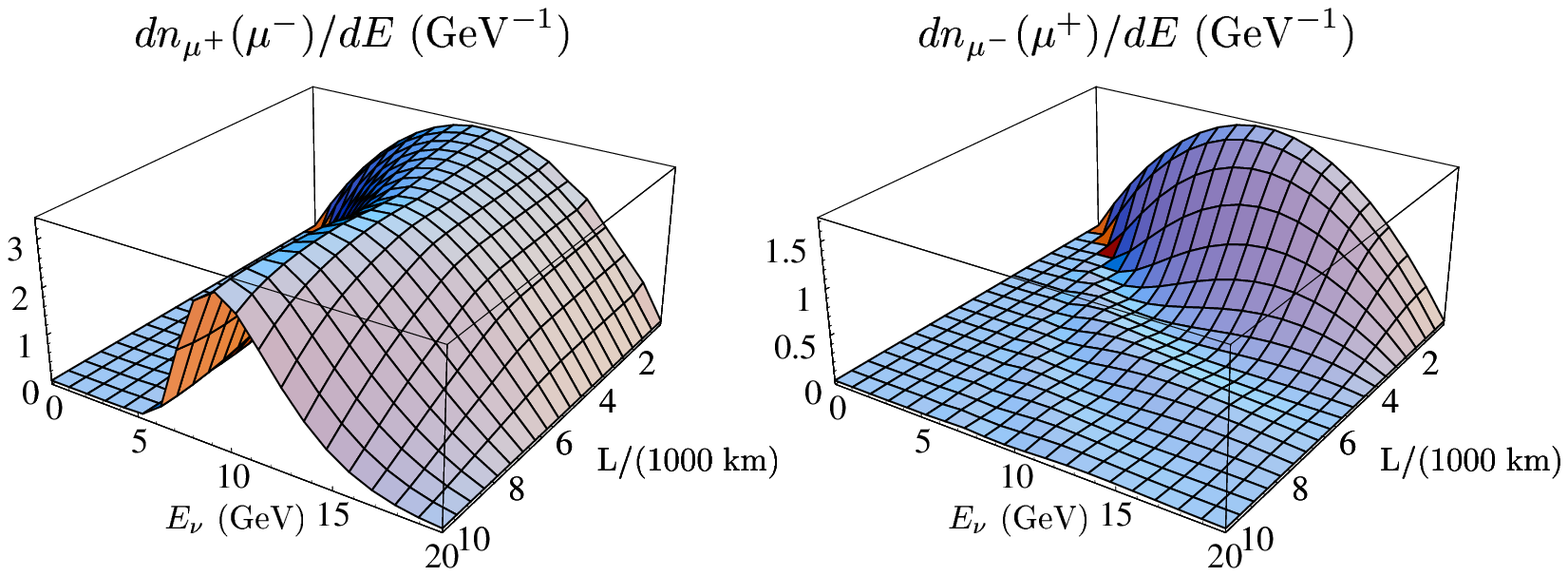,width=1.00\textwidth}
\end{center}
\mycaption{Differential appearance rates for the channels $\reu$ 
(left) and $\reub$ (right) as functions of $L$ and $E$ for the same 
values of the parameters for which fig.~\ref{fig:ratesap} was obtained. 
The asymmetry, i.e. the enhancement of $d\neu /dE$ and the suppression 
of $d\neub /dE$, at $L \gtrsim (5000 - 6000)\mathrm{km}$ is related 
to the MSW effect (see the text). }
\label{fig:rates3D}
\end{figure}
\begin{figure}[hb!]
\begin{center}
\epsfig{file=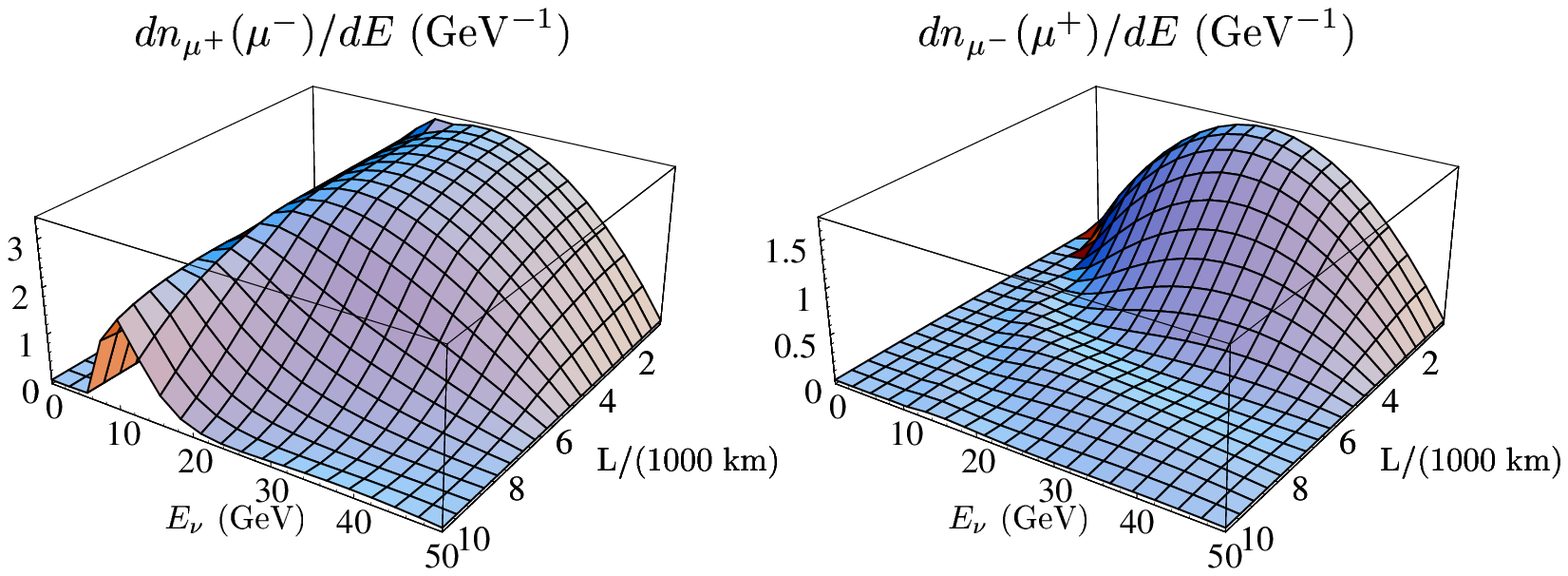,width=1.00\textwidth}
\end{center}
\mycaption{Same as in fig.~\ref{fig:rates3D} but for a beam energy of 
$50\GeV$.}
\label{fig:rates3Db}
\end{figure}

Figs.~\ref{fig:rates3D} and \ref{fig:rates3Db} show in more detail 
the differential event rates of the appearance channels $\reu$ and 
$\reub$ as three dimensional plots over $L$ and $E_\nu$. The figures 
show nicely the enhancement (suppression) due to the MSW mechanism 
in the $\reu$ ($\reub$) channel for large baseline $L$. 
To understand these figures we can look at the values of 
$dn_{\mu^{+}}(\mu^{-})/dE$ and $dn_{\mu^{-}}(\mu^{+})/dE$ 
for three different energies, $E = (5;~10;~15)\GeV$, and 
two baselines $L$. Specifically we compare a relatively 
short baseline $L = L_{0} \equiv 1000\mathrm{~km}$ with a 
long baseline $L=6000\mathrm{~km}$ where matter effects 
are more important. For $L=1000\mathrm{~km}$ ($L=6000\mathrm{~km}$)
the average matter electron density is 
$\bar{N}_e^{man} = 1.45~{\rm cm^{-3}N_{A}}$ 
($\bar{N}_e^{man} \cong 2.0~{\rm cm^{-3}N_{A}}$),
resulting in a resonance neutrino energy of $E_{res} = 15.8\GeV$
($E_{res} = 11.5\GeV$). The resonance energy is thus only
somewhat reduced for larger baselines. The corresponding
values of $\Delta_{31}(L) \equiv L~\dm{31}/(4E)$ and 
the matter effect factors $C_+$ and $C_-$ which enter
in $P^{3\nu}_{E}(\reu)$ and $P^{3\nu}_{E}(\reub)$ 
(see eqs.~(\ref{Ps}), (\ref{PE2nu}), (\ref{Emcos2thetam}) 
and (\ref{cpm})) are displayed in Table~\ref{tab:rates}.

\begin{table}[ht]
\begin{center}
\begin{tabular}{c||c|c|c||c|c|c}
$L$ & \multicolumn{3}{c||}{1000km} & \multicolumn{3}{c}{6000km} \\ \hline
$E$ & $5\GeV$ & $10\GeV$ & $15\GeV$ & $5\GeV$ & $10\GeV$ & $15\GeV$ \\ \hline
\hline
$\Delta_{31}(L)$ & $0.282\pi$ & $0.141\pi $ & $0.094\pi$ & $1.694\pi $ &
$0.847\pi$ & $0.565\pi$ \\ \hline
$C_+$ & $0.686$ & $0.371$ & $0.111$ & $0.568$ & $0.159$ & $0.327$ \\ \hline
$C_-$ & $1.315$ & $1.630$ & $1.945$ & $1.434$ & $1.869$ & $2.304$ \\ \hline
$\Delta_{31}(L)C_+$ & $0.610$ & $0.167$ & $0.031$ & $3.023$ & $0.424$ &
$0.580$\\ \hline
$\Delta_{31}(L)C_-$ & $1.166$ & $0.723$ & $0.575$ & $7.632$ & $4.973$ &
$4.090$ \\ 
\end{tabular}
\caption{Values of $\Delta_{31}(L) \equiv L\dm{31}/(4E)$, 
$C_+$ and $C_-$ for 
$\Delta m^2_{31}=3.5\times 10^{-3}\eV^2$, 
$\sin^22\theta_{23}=1$
and $\sin^2 2\theta_{13}=0.01$ (see the text).}
\label{tab:rates}
\end{center}
\end{table}                     

In general, $\Delta_{31}(L)$ decreases linearly with the increase 
of $E$. At $1000\mathrm{km}$ matter effects are small. $C_{+}$ 
(eq.~\ref{cpm}) decreases as $E$ changes from 5 to 15 GeV, approaching 
the resonance energy from below. Note that due to the matter 
term we have $C_{+} < 1$ and $C_{-} > 1$ for the three  values of $E$ 
of interest. The argument of the oscillating sine factor in 
$P^{3\nu}_{E}(\nu_e \rightarrow \nu_{\mu})$ and in 
$P^{3\nu}_{E}(\bar{\nu}_e \rightarrow \bar{\nu}_{\mu})$
at values of $E$ considered is relatively small allowing an expansion  
of the sine in the oscillation probability 
$P^{3\nu}_{E}(\nu_e \rightarrow \nu_{\mu}) \cong s_{23}^2 
\sin^2(2\theta_{13})~ (\Delta_{31}(L_{0}))^2$. The fact that the 
matter effect factors cancel out reflects the negligibility
of matter effects at short baselines\footnote{Note however that 
for longer baselines where $\Delta \ll 1$ does no longer hold, the 
same conclusion would be wrong.}. 
Neglecting matter effects at short baselines, the maximum of the 
differential event rate spectrum is at one half of the muon beam 
energy which is $10\GeV$ in the case under discussion.

At $L = 6000$ km, one has $\Phi(L) = \Phi(L_{0}) (L_{0}/L)^2 = 
\Phi(L_{0})/36$, where  $\Phi(L)$ is the flux of $\nu_e$ or 
$\bar{\nu}_e$ at distance $L \geq L_{0}$. The argument of the sine 
is large for the anti-neutrino channel giving for the corresponding 
rates the upper bound
\begin{align}
dn_{\mu^{-}}(\mu^{+})/dE \propto& \quad s_{23}^2~g_{\bar{\nu}_e}(x)~ 
{\Phi(L_{0})\over {36}}~
{{\sin^22\theta_{13}}\over{C^2_{-}}} 
\sin^2 \left(\Delta_{31}(L)C_{-}\right) \nonumber \\
\leq & \quad s_{23}^2~g_{\bar{\nu}_e}(x)~ 
{\Phi(L_{0})\over {36}}~{{\sin^22\theta_{13}}\over{C^2_{-}}}~.
\end{align}
Obviously, $dn_{\mu^{-}}(\mu^{+})/dE$ is strongly suppressed 
primarily by the decreasing of the flux. The suppression due 
to the matter effect term ($1/{C_-}^2$) for $E = (5 - 15) \GeV$ 
is by a factor of $\sim (2 - 5)$. The same conclusion is valid 
for $dn_{\mu^{+}}(\mu^{-})/dE$ below the resonance region, say 
at $E = 5$ GeV. 

For neutrinos in the resonance region $E \cong (10 - 15)\GeV$, the 
argument of the sine is small even for very long baselines.
Like in the case of short baselines an expansion becomes possible:
\begin{align}
dn_{\mu^{+}}(\mu^{-})/dE \propto& \quad
s_{23}^2~g_{\nu_e}(x)~
\Phi(L_{0}) ({L_{0} \over L})^2~
{{\sin^22\theta_{13}}\over{C^2_{+}}} \sin^2 (\Delta_{31}(L)C_{+}) \nonumber\\
\cong& \quad s_{23}^2~g_{\nu_e}(x)~\Phi(L_{0})~ \sin^2(2\theta_{13})~
({\Delta m^2_{31} \over{4E}}L_{0})^2~.
\end{align} 
This is approximately equal to the same differential event 
rate at short baselines. Thus, in the resonance region 
($E \cong (10 - 15)\GeV$) and at sufficiently long baselines 
($L \gtap 4000$), matter effects lead to a strong suppression 
of anti-neutrino event rate while keeping the neutrino event rate 
essentially constant with the change of $L$ from 
$\sim 1000\mathrm{km}$ to $\sim 7000\mathrm{km}$ 
(see fig.~\ref{fig:rates3D}). In vacuum, however, the neutrino 
rate as well as the anti-neutrino rate would show the same strong 
$(L_{0}/L)^2$ suppression at long baselines.

For fixed $L \gtrsim 4500$ km, say at $L = 6000$ km, the matter
effects lead to a strong enhancement of $dn_{\mu^{+}}(\mu^{-})/dE$ 
as a function of $E$ in a region which is somewhat wider than
the MSW resonance region, $E \sim (8 - 15)$ GeV, in spite of the 
fact that in this region $\Delta_{31}(L)C_{+} < 0.6$ and 
\begin{align}
dn_{\mu^{+}}(\mu^{-})/dE \propto& \quad s_{23}^2~g_{\nu_e}(x)~\Phi(L)
{{\sin^22\theta_{13}}\over{C^2_{+}}} \sin^2 (\Delta_{31}(L)C_{+}) \nonumber \\
\cong& \quad s_{23}^2~g_{\nu_e}(x)~
\Phi(L)~\sin^22\theta_{13}~(\Delta_{31}(L))^2.
\end{align}
We have in the indicated energy region $\Delta_{31}(L) > 1$ and 
the enhancement of $dn_{\mu^{+}}(\mu^{-})/dE$ compared to 
the case of $\nu_e \leftrightarrow \nu_{\mu}$ vacuum-oscillations
is actually given by the ratio of $(\Delta_{31}(L))^2$ and 
$\sin^2 (\Delta_{31}(L))$. 

Fig.~\ref{fig:rates3Db} shows the results corresponding to 
fig.~\ref{fig:rates3D} for a beam energy of $50\GeV$. At very 
long baselines, the neutrino rate spectrum clearly 
peaks at the energy where the MSW-resonance condition is fulfilled. 
The change of the shape of the spectrum with the baseline $L$ is 
more pronounced now because the resonance energy does not coincide 
with the maximum of the beam spectrum as it approximately does for a 
beam energy of $20\GeV$.  The baseline and muon energy 
dependence of matter effects will be further discussed in the next
section in connection with a quantitative analysis of the significance
of the effects shown in figs.~\ref{fig:ratesap} and \ref{fig:ratesdis}.

%%%%%%%%%%%%%%%%%%%%%%%%%%%%%%%%%%%%%%%%%%%%%%%%%%%%%%%%%%%%%%%%%%%%%
%%%%            SECTION: Significance of matter effects          %%%%
%%%%%%%%%%%%%%%%%%%%%%%%%%%%%%%%%%%%%%%%%%%%%%%%%%%%%%%%%%%%%%%%%%%%%

\section{Statistical Significance of Matter Effects}

We have seen in the previous section that matter effects change 
the total event rates in the appearance channels $\reu$ and
$\reub$ in very long baseline experiments in a drastic way. 
Such experiments would therefore offer unique possibilities 
to observe matter effects and to test the predictions of the 
MSW theory. In order to study the capabilities of a neutrino factory
experiment quantitatively,  we must first define the meaning 
of ``observing matter effects''. One of the most interesting 
possibilities would be a detailed observation of the shape of 
the neutrino energy spectrum which is modified by the MSW effect 
in a very characteristic way. This would allow to 
test non-trivial predictions of the MSW theory and would 
allow to unambiguously attribute the enhancement/depletion of 
the total number of neutrino events to matter effects. 
High differential event rates and a good 
calibration of the detector would however be necessary for 
this option. We will discuss this possibility in
sect.~\ref{sec:spectrum}. In this section we confine ourselves to a
discussion of the significance of MSW effects in total event rates. 

Matter effects produce deviations of the total number of wrong 
sign muon events $\neu$, $\neub$ from what is expected in the 
absence of matter. The discussion above clearly 
shows that such deviations occur in opposite directions in the 
appearance channels. Suppose that a certain number of wrong sign 
muon events in the neutrino (antineutrino) appearance channel 
$\neu$ ($\neub$) is measured, while $\VeV{\neuvac}$ ($\VeV{\neubvac}$) 
would be expected in the absence of matter effects. We want to 
determine the confidence level at which $\neu$ and $\neub$ could 
represent statistical fluctuations around the expected values in 
vacuum $\VeV{\neuvac}$, $\VeV{\neubvac}$. We follow the procedure 
proposed by the Particle Data Book \cite{PDG99} and calculate 
confidence levels by using
\begin{equation}
\label{chi}
\begin{split}
\chi^2 = 2\, &\left[\VeV{\neuvac}-\neu\right]
+2\neu\log\frac{\neu}{\VeV{\neuvac}} \\
+ 2\, &\left[\VeV{\neubvac}-\neub\right]
+2\neub\log\frac{\neub}{\VeV{\neubvac}} \quad .
\end{split}
\end{equation}

\begin{figure}[b!]
\begin{center}
\epsfig{file=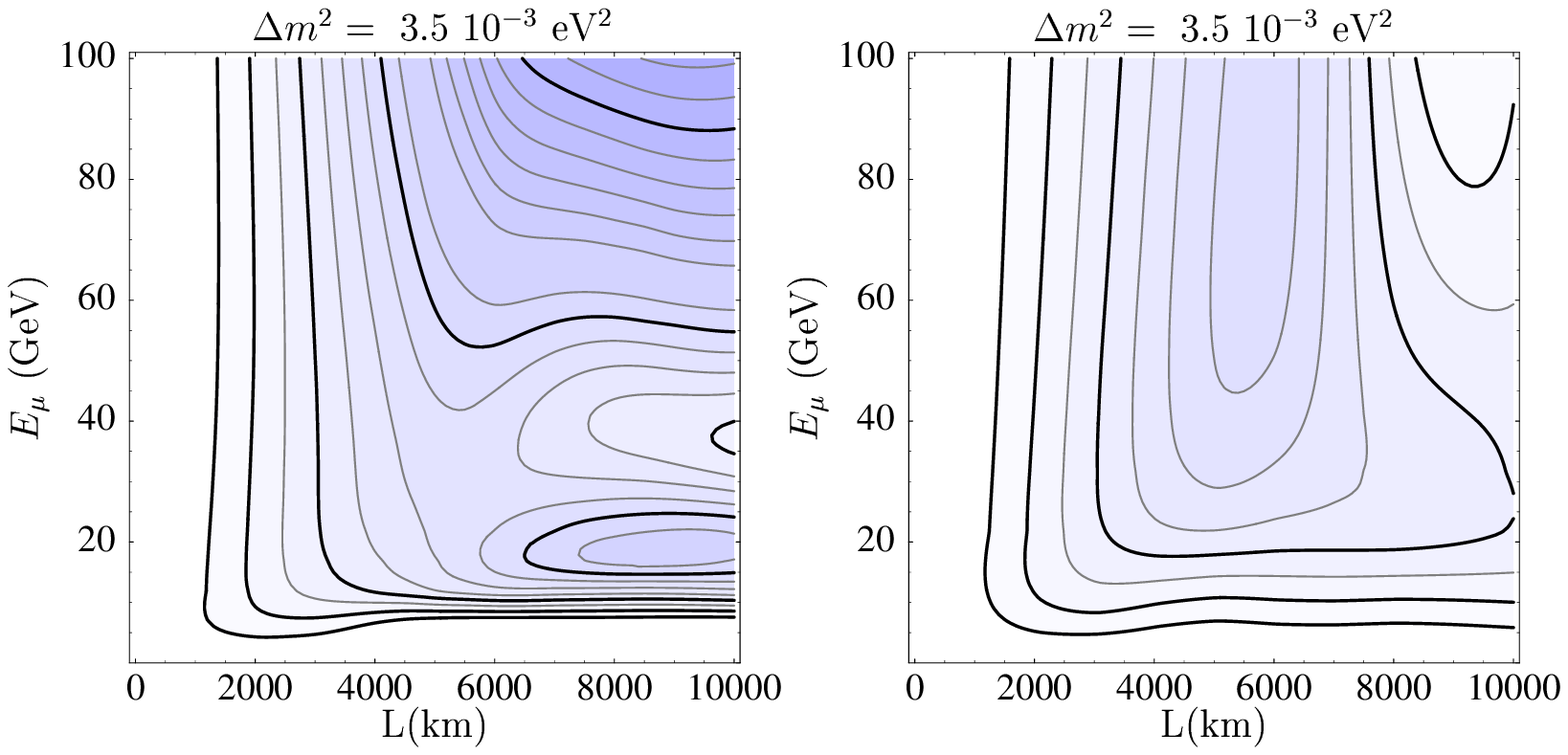,width=\textwidth}
\end{center}
\mycaption{Contour lines of $n_\sigma$ in the $L$-$E_\mu$ plane. 
We use $\dm{31}=3.5\cdot 10^{-3}\eV^2$ and 
$N_\mu \NKT\epsilon=10^{21}$ and the solid contour lines correspond 
to $n_\sigma = 10 \sin 2\theta_{13} \cdot \{1, 2, 4, 8, 16\}$. The 
left plot assumes that $\theta_{13}$ is known, while the right plot 
is obtained by varying $\theta_{13}$ in the range currently allowed.}
\label{fig:opt}
\end{figure}

\begin{figure}[ht!]
\begin{center}
\epsfig{file=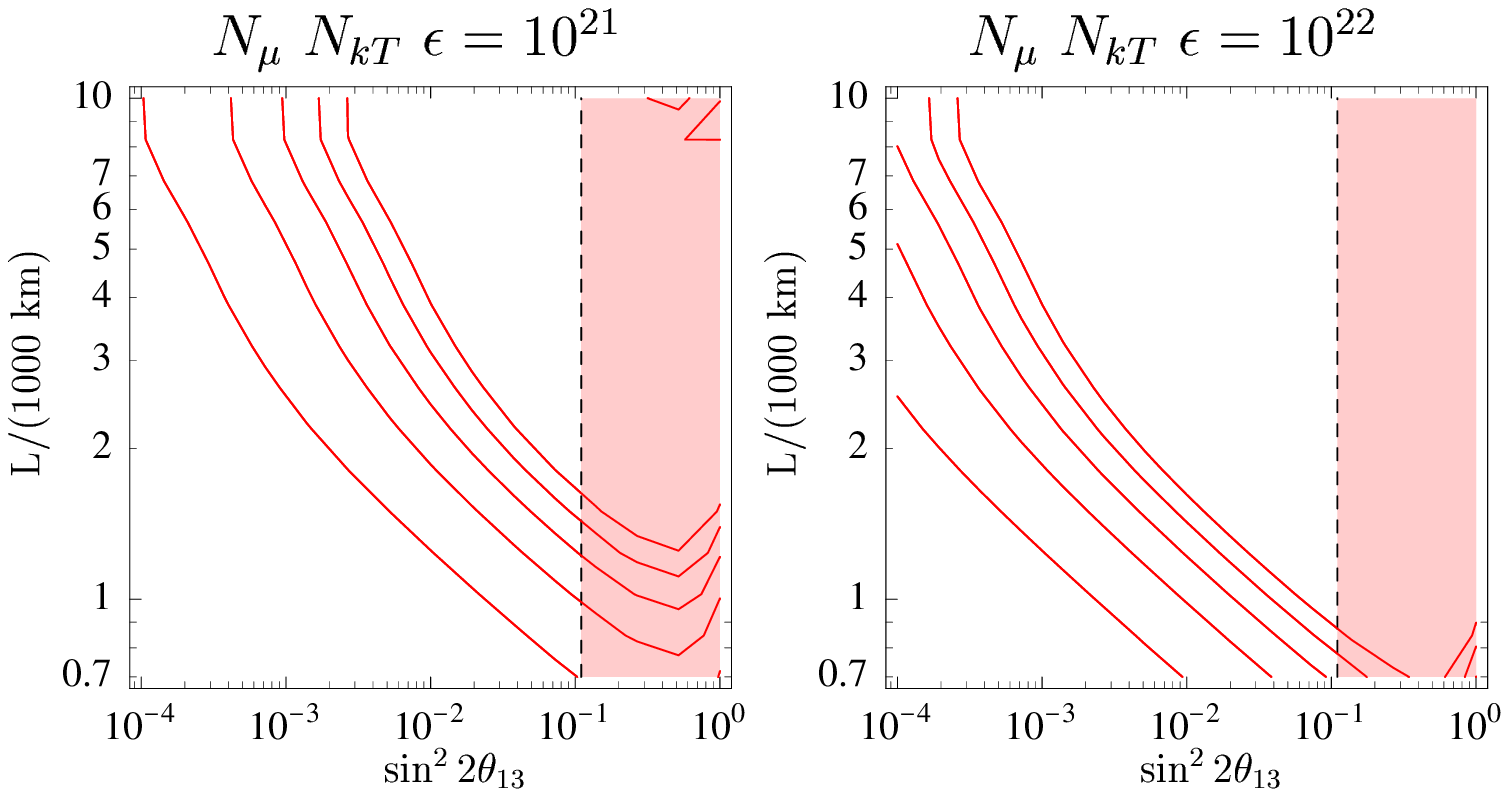,width=1.00\textwidth}
\end{center}
\mycaption{Contour lines of $n_\sigma$ corresponding to
$n_\sigma=1,2,3,4,5$ in the $\sin^22\theta_{13}$-$L$ plane 
for $E_\mu=20\GeV$ and the two different values 
$N_\mu \NKT\epsilon = 10^{21}$ (left plot) and 
$N_\mu \NKT\epsilon = 10^{22}$ (right plot).}
\label{fig:CL20}
\end{figure}

\begin{figure}[ht!]
\vspace*{6mm}
\begin{center}
\epsfig{file=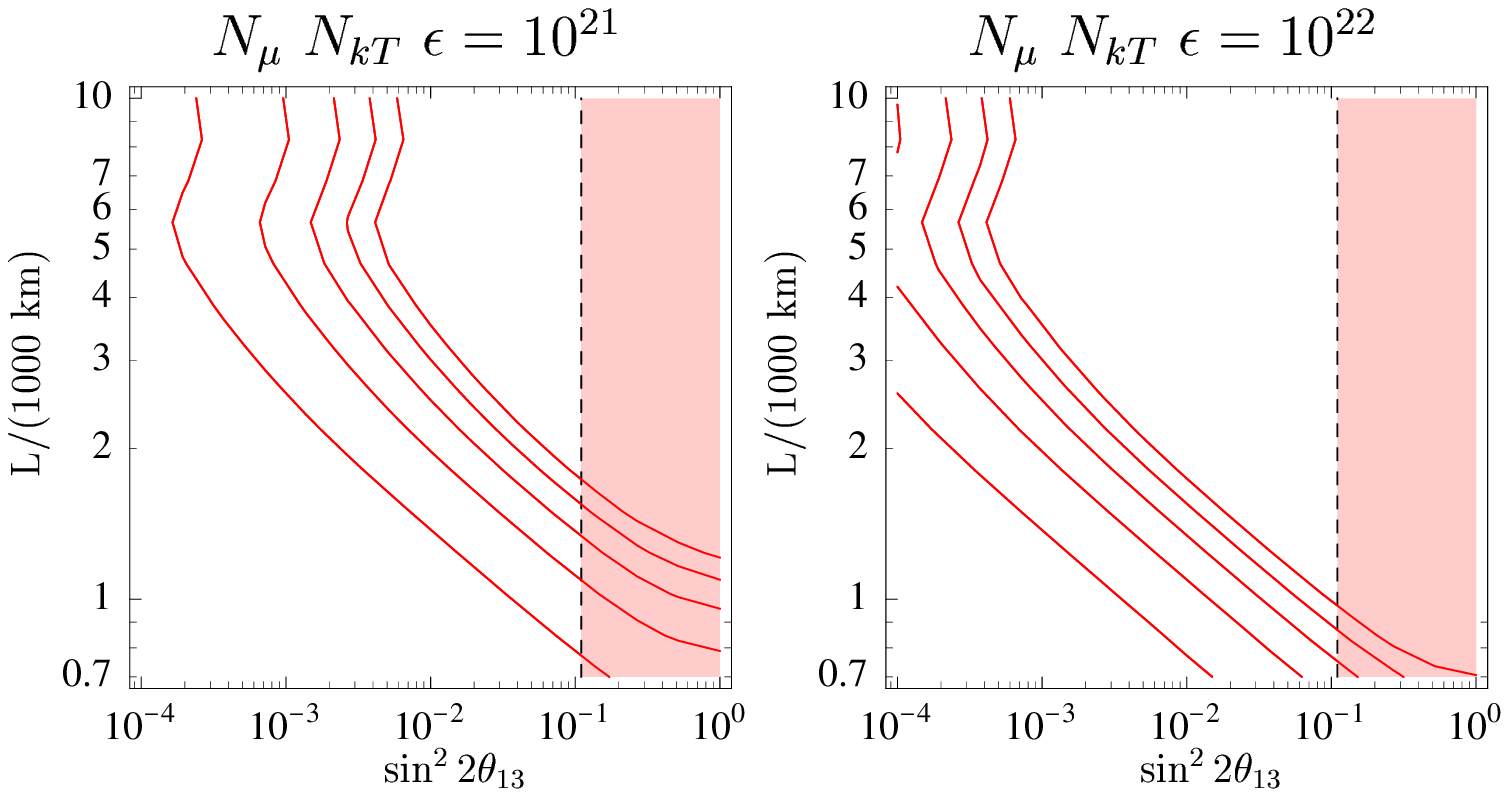,width=1.00\textwidth}
\end{center}
\mycaption{Same as in fig.~\ref{fig:CL20} but for $E_\mu=50\GeV$.}
\label{fig:CL50}
\end{figure}

The corresponding ``number of standard deviations'' is given by
$n_\sigma\equiv\sqrt{\chi^2}$. This prescription incorporates the
available information in both the measured numbers $\neu$, $\neub$ in
the most complete way. Note that the results which will be shown later 
with this method in figures~\ref{fig:opt}-\ref{fig:CL50} below assume 
that $\theta_{13}$ is known with some precision before such an analysis 
is performed. This could e.g. be realized by a measurement of the 
$\reu$ and/or $\reub$ rates at a relatively short baseline, say, below 
1000~km. If $\theta_{13}$ were only limited from above then one would 
have to vary $\theta_{13}$ in the vacuum rates entering the definition 
of $\chi^2$ in eq.~(\ref{chi}) in the allowed range and find the 
minimum. This would reduce the significance at largest baselines. 
The statistical method defined above in eq.~(\ref{chi}) 
can be understood intuitively by defining an asymmetry 
\begin{equation}
A:=\frac{\Delta}{\Sigma} = \frac{\neu-2\neub}{\neu+2\neub}
\end{equation}
which is obviously very sensitive to effects which affect $\neu$ 
and $\neub$ in a opposite way, 
while at the same time a number of common systematic effects drop 
out. For small $\Sigma^\mathrm{vac}$ (i.e. small beam energies) 
this method is related to comparing the absolute asymmetry expected 
in matter with the expected fluctuations of this quantity in the 
vacuum-case $\chi = \Delta / \delta \Delta^\mathrm{vac}$.
For large $\Sigma^\mathrm{vac}$ this is equivalent to doing the same 
with the asymmetry: $\chi = A/\delta A^\mathrm{vac}$.
Note that the sign of $\dm{31}$ can be determined from $A$ or 
equivalently from $\Delta$. This can easily be seen by observing, 
for example, that for $N_{\mu^+}=N_{\mu^-}$ the difference 
$P^{3\nu}_E(\reu)- P^{3\nu}_E(\reub)$, with $P^{3\nu}_E(\reu)$ 
and $P^{3\nu}_E(\reub)$ given by eq.~(\ref{pro}), enters in 
$\Delta$. The sign of $\Delta$ and $A$ depends therefore 
unambiguously on whether $C_+$ or $C_-$ is bigger 
than one (and the other smaller than one). 
Via the definition of $C_\pm$ given in eq.~(\ref{cpm})
this is in turn unambiguously related to the sign of $\dm{31}$.

Before we show the numerical results, let us discuss the qualitative
dependence of $n_\sigma$ on the relevant parameters. The
dependence of $n_\sigma$ on the intensity of the muon source and the
detector size and efficiency, as well as the dependence on the mixing
parameters follows simply from the previous section: 
\begin{equation}
n_\sigma\propto
(N_\mu\NKT\epsilon_\mu)^{1/2} \quad\text{and}\quad
n_\sigma\propto \sin\theta_{23}\sin2\theta_{13}~.
\end{equation}
The dependence of $n_\sigma$ on the baseline $L$ and the muon 
energy $E_\mu$ is less trivial. In general, matter effects increase 
with the baseline $L$. Thus a long baseline is essential 
for the observation of matter effects. For relatively small baselines 
$L \le 730\,\text{km}$, in fact, matter effects mostly cancel in 
eq.~(\ref{pro}), thus making such baselines better suited for 
CP-violation measurements~\cite{gavela-c,romanino}. The dependence on 
$E_\mu$ needs more explanation and there is also some dependence on the 
statistical method used. In our simple approach which takes into 
account only total rates, the significance to matter effects grows
first strongly with the increase of the beam energy. At very 
long baselines, where matter effects are non negligible, a maximum is 
reached at a beam energy which allows a maximal number of neutrinos 
to fall into the MSW-resonance energy region. After a 
small depletion the significance starts to grow again. This is due
to the fact that the rates of high energy ($\gtrsim 50\GeV$) neutrinos and 
anti-neutrinos are strongly suppressed due to the matter effects. 
For large beam energies it is, as was already mentioned earlier,
crucial to know the precise value of $\sin^2 2\theta_{13}$. Otherwise 
it is difficult to distinguish matter suppression from a reduced event 
rate due to a smaller $\theta_{13}$ in absence of matter effects (see 
fig.~\ref{fig:ratesap} for large L). 

The dependence of $n_\sigma$ on $L$ and $E_\mu$ is shown in
fig.~\ref{fig:opt} for $\dm{31}=3.5\cdot 10^{-3}\eV^2$ and 
$N_\mu \NKT\epsilon=10^{21}$, where the contour lines of $n_\sigma$ 
are plotted in the $L$-$E_\mu$ plane. The solid contour lines correspond 
to $n_\sigma = 10 \sin 2\theta_{13} \cdot \{1, 2, 4, 8, 16\}$.
The left plot assumes that $\theta_{13}$ is 
known, while the right plot is obtained by varying $\theta_{13}$ 
in the range currently allowed. With the simple "total
rates based" method described above and assuming in the following 
discussion that $\theta_{13}$ is known, any $L$ from the interval 
$(4.0 - 10.0)\times 10^{3}$ km would be suitable for the purpose if $\Delta
m^2_{31} = 3.5\times 10^{-3}\eV^2$, $\sin^22\theta_{13} = 0.01$ and
$E_{\mu} = 20$ GeV, with the sensitivity to matter effects increasing
with $L$. For $E_{\mu} = 50$ GeV, the sensitivity varies very little
for $L \cong (4.0 - 10.0)\times 10^{3}$ km. 
Let us note that at $L \gtrsim 3.0\times 10^{3}$ km the matter effects
are substantial for any neutrino energy $E$ either in the $\nu_e
\rightarrow \nu_{\mu}$ and/or in the $\bar{\nu}_e \rightarrow
\bar{\nu}_{\mu}$ channel, depending on the sign of $\Delta m^2_{31}$:
at $L \gtrsim 3.0\times 10^{3}$ km we have
$\sqrt{2}G_{F}\bar{N}_e^{man}L/2 \gtrsim 0.95$. The matter effects 
can be negligible for certain intervals of neutrino energies
for $L \lesssim 2.0\times 10^{3}$ km. For relatively large $E$ and  
$L \lesssim 2.0\times 10^{3}$ km, the oscillating factors in 
$P^{3\nu}_{E}(\nu_e \rightarrow \nu_{\mu})$ and 
$P^{3\nu}_{E}(\bar{\nu}_e \rightarrow \bar{\nu}_{\mu})$,
eqs.~(\ref{Ps}), (\ref{PE2nu}), (\ref{Emcos2thetam}) and (\ref{cpm}),
can be expanded in power series of their arguments and 
matter effects cancel in the leading orders of these expansions.
Moreover, one has $\sin^2(\Delta_{31}(L)) \cong (\Delta_{31}(L))^2$.
At $L = 10^{3}$ km and for
$\Delta m^2_{31} = 3.5\times 10^{-3}\eV^2$, $\sin^22\theta_{13} = 0.01$,
for instance, the earth matter effects cause a
difference between $P^{3\nu}_{E}(\nu_e \rightarrow \nu_{\mu})$, eq.
(\ref{Ps}), and the probability of the $\nu_e \rightarrow \nu_{\mu}$
transitions in vacuum $P^{3\nu}_{vac}(\nu_e \rightarrow \nu_{\mu})$,
eq.~(\ref{Pmu2tau}), which does not exceed $\sim 5\times 10^{-5}$ at $E \geq
10$ GeV. We have, however,  $|P^{3\nu}_{E}(\bar{\nu}_e \rightarrow
\bar{\nu}_{\mu}) - P^{3\nu}_{vac}(\bar{\nu}_e \rightarrow
\bar{\nu}_{\mu})| \leq  10^{-4}~(10^{-5})$ for $E \geq 10~(20)$
GeV. For $\Delta m^2_{31} = 7.0\times 10^{-3}\eV^2$, the same
conclusions hold for approximately two times higher neutrino
energies. Note that $P^{3\nu}_{vac}(\nu_e \rightarrow \nu_{\mu}) \sim 
s^2_{23}\sin^22\theta_{13}$ and for the values of $\sin^22\theta_{13}$
and $s^2_{23}$ used in the above examples 
we have $s^2_{23}\sin^22\theta_{13} = 0.005$. 
Thus, baselines shorter than about $2000\,\text{km}$ and parent muon
energies $E_{\mu} \gtrsim (20 - 25)$ GeV seem to be better suited for
searches of CP-violation in neutrino oscillations whose source is the
lepton mixing matrix.  The magnitude of the asymmetry between the
probabilities due to matter effects effectively grows quadratically
with the baseline until the argument $\Delta_{31}(L)C_{+}$ and/or
$\Delta_{31}(L)C_{-}$ approaches and exceeds one~\cite{romanino}.

The significance of the effect, i.e. the number of standard deviations 
$n_\sigma$, depends most crucially on $\theta_{13}$ and $L$. We
illustrate this dependence in fig.~\ref{fig:CL20} again for 
$\dm{31}=3.5\cdot 10^{-3}\eV^2$, where the 
contour lines corresponding to $n_\sigma=1,2,3,4,5$ are plotted 
in the $\sin^22\theta_{13}$-$L$ plane for two values of the 
product $N_\mu \NKT\epsilon$: $10^{21}$ (left plot) and 
$10^{22}$ (right plot). The muon energy is in both cases 
of fig.~\ref{fig:CL20} $E_\mu=20\GeV$, while fig.~\ref{fig:CL50}
shows the same plots with identical parameters for $E_\mu=50\GeV$. 
The vertical dashed lines represent in all these figures the upper 
limit on $\sin^22\theta_{13}$. Figs.~\ref{fig:CL20} and~\ref{fig:CL50} 
show that matter effects could be observed in the total event rates 
for given baseline $L$ in a rather large $\sin^22\theta_{13}$ interval,
while non-observation implies very strong upper bounds on 
$\sin^22\theta_{13}$. Figs.~\ref{fig:CL20} and~\ref{fig:CL50} show in 
other words the $\sin^22\theta_{13}$ range where the enhancement/depletion 
of the total appearance rates in vacuum due to matter effects is 
statistically significant for a given confidence level. In those ranges 
one can not only observe the deviations from the results expected 
in vacuum, but also the deviation from the results which one 
would get in matter if the sign of $\dm{31}$ were reverted.
A measurement of the sign of $\dm{31}$ would therefore be
possible in the $\sin^2 2\theta_{13}$ range where matter 
effects are statistically significant at a given confidence level.

\begin{figure}[ht!]
\begin{center}
\epsfig{file=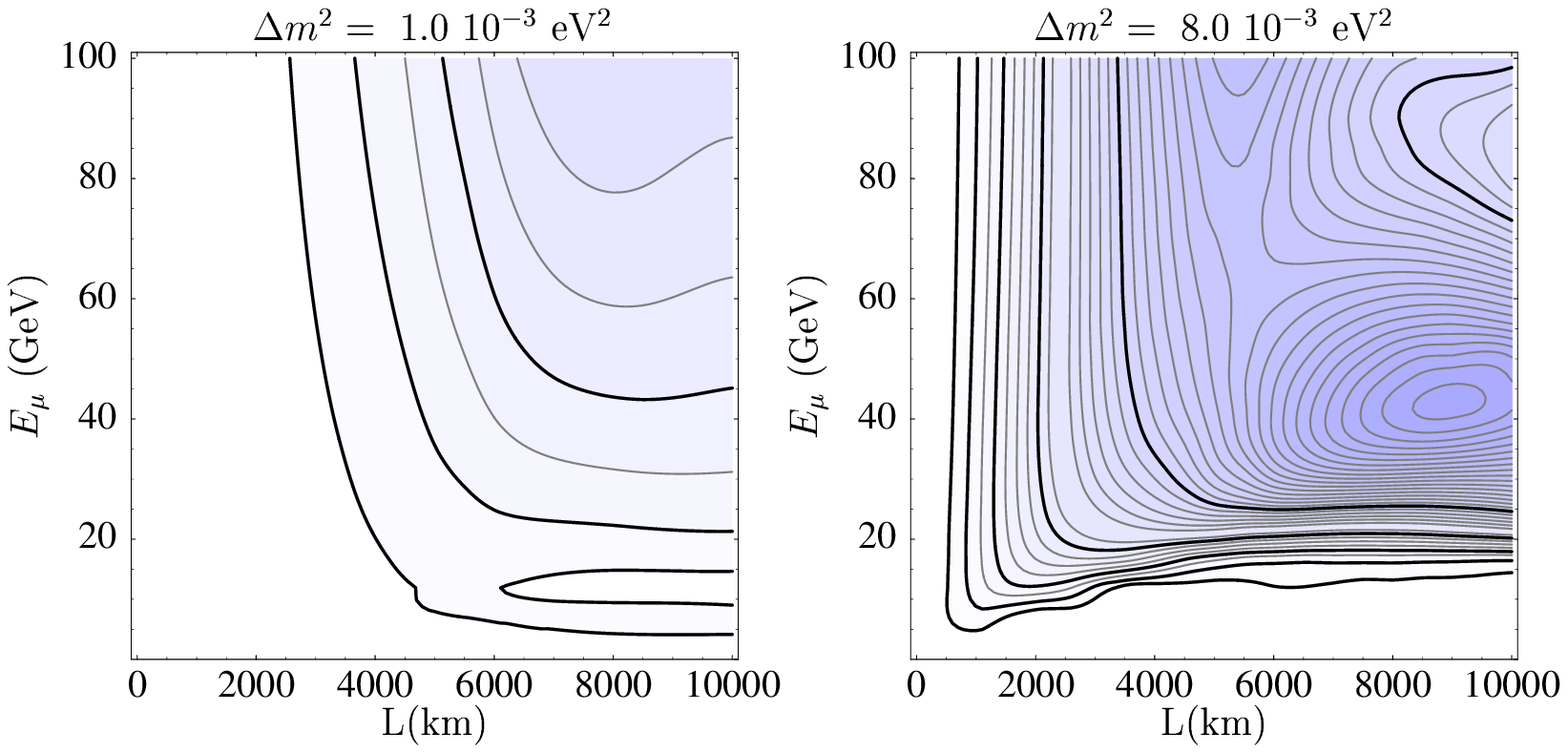,width=1.00\textwidth}
\end{center}
\mycaption{Sensitivity of the statistical significance of
matter effects to the value of $\dm{31}$ analogous to 
fig.~\ref{fig:opt}. The left plot uses 
$\dm{31}= 1.0\cdot 10^{-3}\eV^2$ while for the right 
plot $\dm{31}= 8.0\cdot 10^{-3}\eV^2$ with otherwise
unchanged parameters. As in fig.~\ref{fig:opt} the solid 
contour lines correspond to $n_\sigma = 10 \sin 2\theta_{13}
\cdot \{1, 2, 4, 8, 16\}$.}
\label{fig:optdm31}
\end{figure}

\begin{figure}[t!]
\vspace*{14mm}
\begin{center}
\epsfig{file=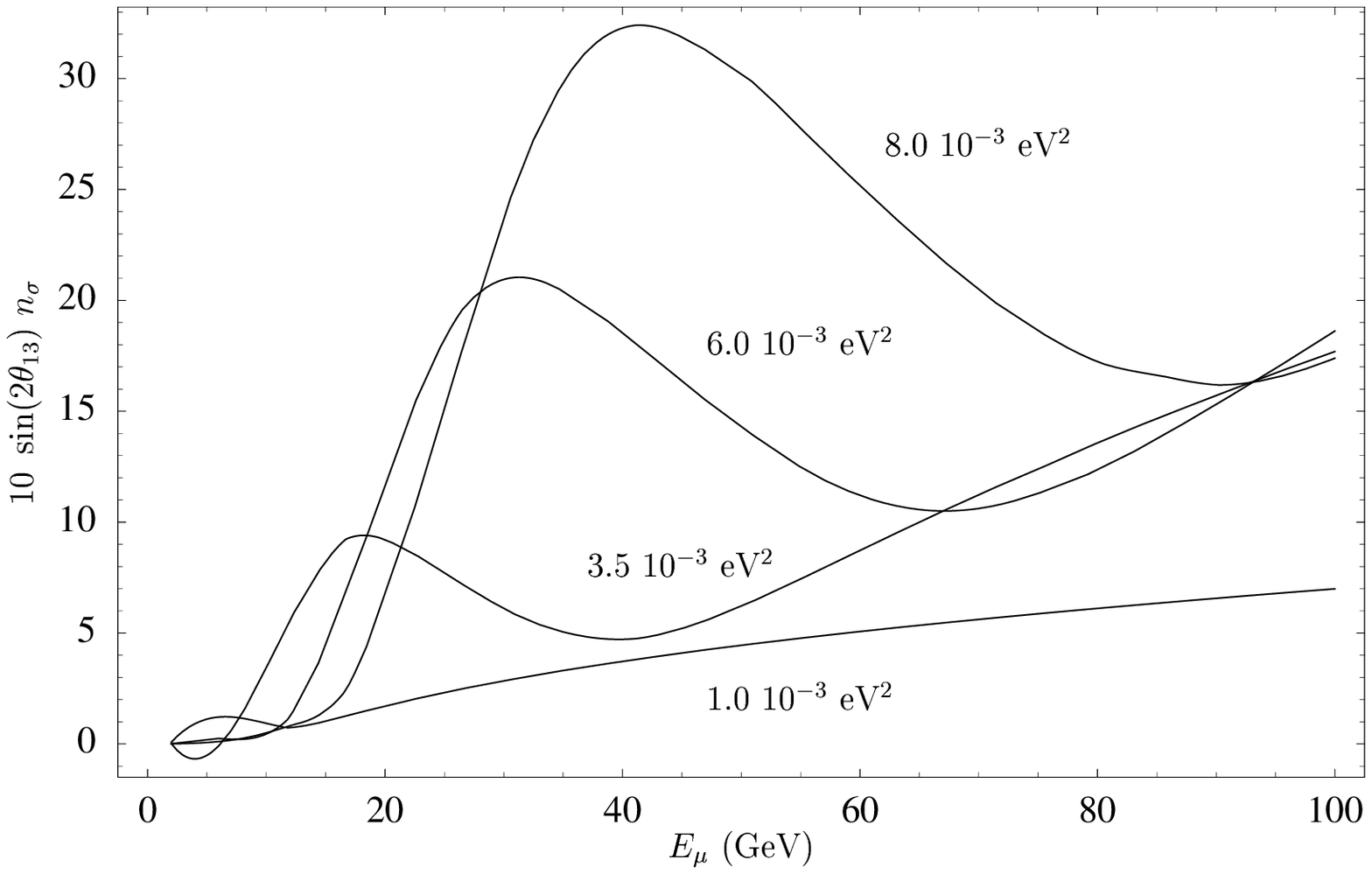,width=0.65\textwidth}
\end{center}
\mycaption{Sensitivity of the statistical significance of matter 
effects to the value of $\dm{31}$ for fixed $L=8000$~{\rm km} as a function 
of $E_\mu$. The lines show $10 \sin 2\theta_{13} n_\sigma $ for 
different $\dm{31}$ values.}
\label{fig:optdm31cut}
\end{figure}

Finally we show in figs.~\ref{fig:optdm31} and~\ref{fig:optdm31cut}
the sensitivity of the statistical significance to the chosen
value of $\dm{31}$. Fig.~\ref{fig:optdm31} is exactly the
same plot as fig.~\ref{fig:opt} with minimal and maximal
$\dm{31}$. In the left plot we have $\dm{31}= 1.0\cdot 10^{-3}\eV^2$ 
and in the right plot $\dm{31}= 8.0\cdot 10^{-3}\eV^2$.
Fig.~\ref{fig:optdm31cut} shows the statistical significance
(i.e. $10 \sin 2\theta_{13} n_\sigma $) for different $\dm{31}$ 
values for $L=8000$~km as a function of $E_\mu$. For very long 
baselines, one can see from fig.~\ref{fig:optdm31cut} that 
increasing $\dm{31}$ mainly shifts the resonance energy to higher 
values thus demanding higher beam energies to reach optimal
statistical significance. The local maximum in fig.~\ref{fig:optdm31}
shifts to $10\GeV$ for minimal $\dm{31}$ and to $40\GeV$ for maximal
$\dm{31}$. Improved knowledge of $\dm{31}$ would thus in principle 
allow to discuss optimization issues, but such a study should also 
include systematics and backgrounds. Moreover, the energy distribution 
of wrong sign muon events would add important information to the simple 
counting of events. A detailed discussion of muon energy optimization 
depends therefore on the way that information will be exploited.

%%%%%%%%%%%%%%%%%%%%%%%%%%%%%%%%%%%%%%%%%%%%%%%%%%%%%%%%%%%%%%%%%%%%%
%%%%    SECTION: \Delta m_12^2 \neq 0                            %%%%
%%%%%%%%%%%%%%%%%%%%%%%%%%%%%%%%%%%%%%%%%%%%%%%%%%%%%%%%%%%%%%%%%%%%%

\section{Subleading $\dm{21}$-effects}
\label{sec:sun}

\begin{figure}[ht!]
\begin{center}
\epsfig{file=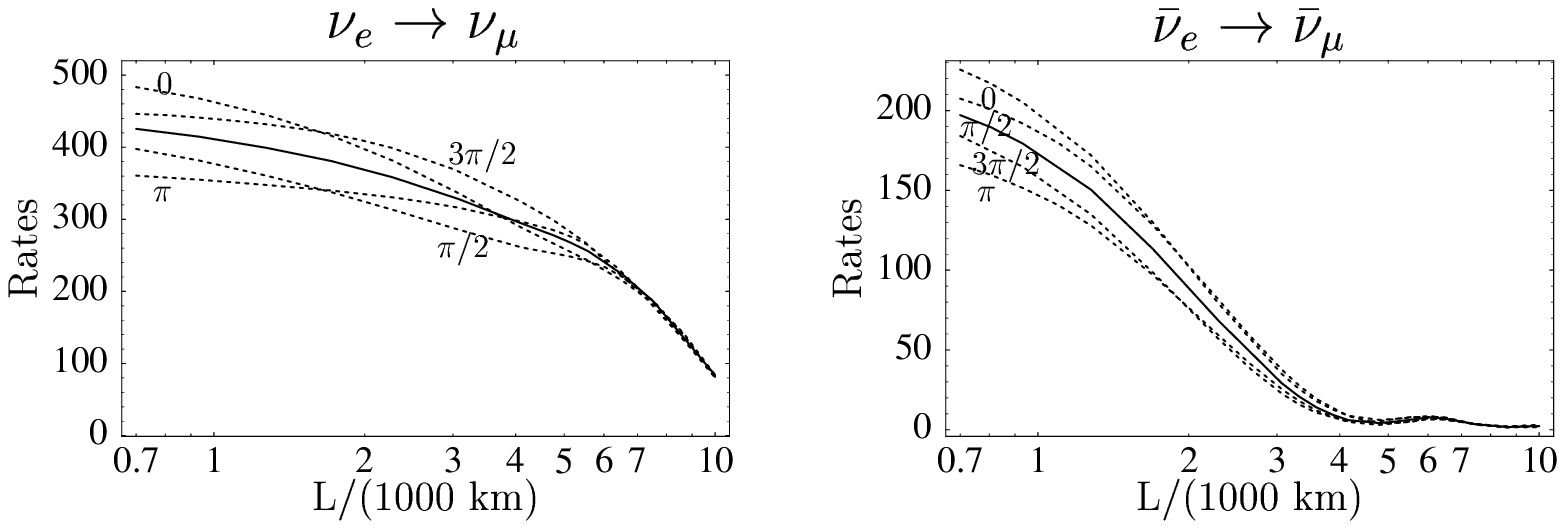,width=1.00\textwidth}
\end{center}
\mycaption{Appearance event rates $\neu$, $\neub$ in matter with
subleading $\dm{21} = 10^{-4}\eV^2$ and four possible values of the
CP-phase $\delta = 0,\;\pi/2,\;\pi,\;3\pi/2$ against baseline $L$ 
(dashed lines) compared with the corresponding event rates with 
negligible $\dm{21}$ (solid lines) for the channels $\reu$ (left) 
and $\reub$ (right). Both figures correspond to 
$N_\mu=2\cdot 10^{20}$, $\epsilon=50\%$, $E_\mu=20\GeV$, 
$\sin^2 2\theta_{23}=1$, $\sin^2 2\theta_{12}=0.8$ and 
$\sin^2 2\theta_{13}=0.1$.}
\label{fig:sun1}
\end{figure}

\begin{figure}[ht!]
\begin{center}
\epsfig{file=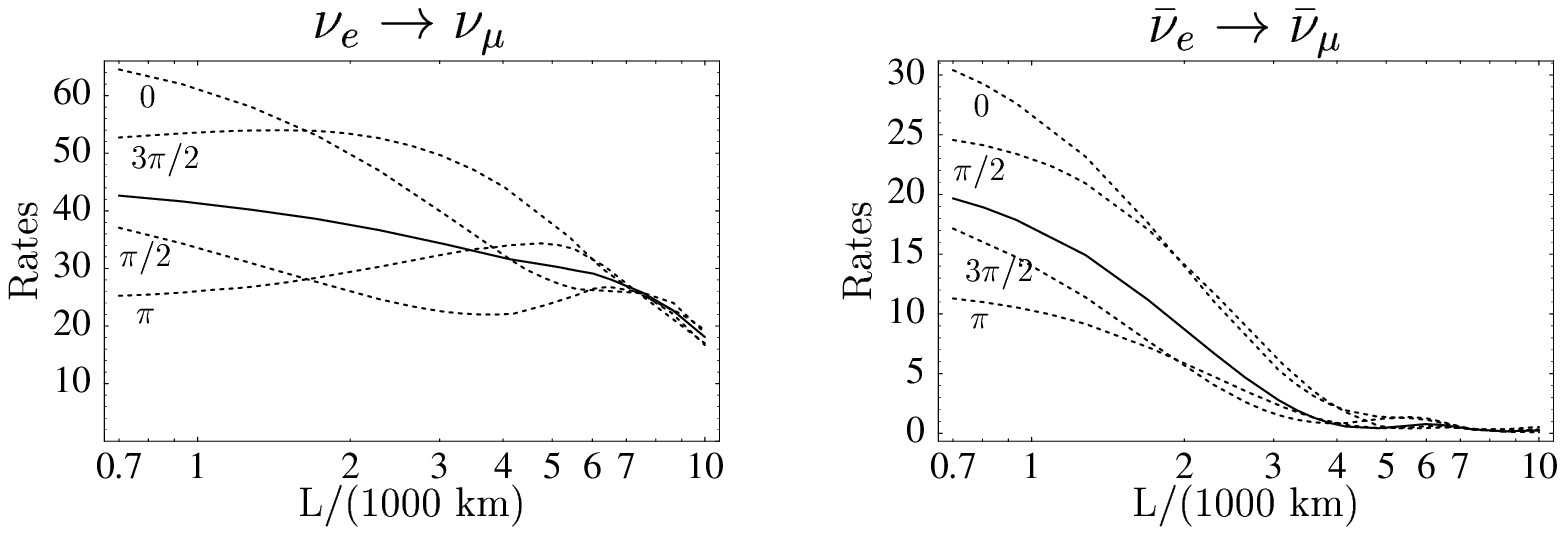,width=1.00\textwidth}
\end{center}
\mycaption{Same as fig.~\ref{fig:sun1} but for $\sin^2 2\theta_{13}=0.01$.}
\label{fig:sun2}
\end{figure}

The results shown so far were obtained in the limit $\dm{21}=0$ 
which is, as already explained, a perfect approximation for 
$\dm{21} \ll 10^{-4}\eV^2$ and/or $\sin^2 2\theta_{12}\ll 1$, 
i.e. for the VO and SMA~MSW solutions of the solar neutrino problem. 
The LMA~MSW solution, that we will consider in this section, 
allows however $\dm{21}$ values up to $2\cdot 10^{-4}\eV^2$ 
and prefers $\sin^2 2\theta_{12}\simeq 0.8$~\cite{concha}, 
so that effects associated to $\dm{21}$ can become important,
especially for the largest $\dm{21}$ values in the LMA range. 
Two more mixing parameters, namely $\theta_{12}$ and $\delta$,
become relevant if $\dm{21}$ is non-negligible 
\cite{KP3nu88,KuoP87} (see also, e.g., \cite{sato1}), as the 
expression for  $P^{3\nu}_{E}(\nu_{e} \rightarrow \nu_{\mu})$,
eq.~(\ref{Pemudelta}), including the leading order
$\dm{21}-$corrections shows. While $\theta_{12}$ is rather 
constrained, so that we will use $\sin^2 2\theta_{12}=0.8$ in 
the following numerical results, any value of $\delta$ in its 
range $0\leq\delta< 2\pi$ is allowed at present. As it is clear 
from eq.~(\ref{Pemudelta}), in order to calculate the effects 
associated with a non-negligible $\dm{21}$, a value of
$\delta$ must be specified. In fig.~\ref{fig:sun1}, the total rates 
in the appearance channels $\reu$ (left) and $\reub$ (right) for
$\dm{21}=0$ (solid line) are compared with the total rates for
$\dm{21}=10^{-4}\eV^2$ and four possible values of $\delta$ in its
range, $\delta = 0,\;\pi/2,\;\pi,\;3\pi/2$ (dashed lines). 
As it follows from eq.~(\ref{Pemudelta}) and these figures,
the size of the effects depends crucially on the value of $\theta_{13}$. 
Fig.~\ref{fig:sun1} assumes a value of $\theta_{13}$ at its upper 
limit, i.e. $\sin^2 2\theta_{13}=0.1$, whereas fig.~\ref{fig:sun2}
shows the effects for a $\sin^2 2\theta_{13}$ one order of magnitude
smaller, $\sin^2 2\theta_{13}=0.01$. Both figures were obtained for 
$N_\mu \NKT\epsilon=10^{21}$.

Figs.~\ref{fig:sun1} and~\ref{fig:sun2} illustrate several
interesting features of the $\dm{21}$ effects. First of all, a
comparison of figs.~\ref{fig:sun1} and~\ref{fig:sun2} confirms 
that the relative size of the effects grows when $\theta_{13}$ gets
smaller~\cite{DFLR}. This is because the zeroth order approximations
(in $\dm{21}$) for the appearance probabilities have a
$\sin^22\theta_{13}$ suppression, whereas the linear $\dm{21}$
corrections (CP-conserving and violating) are suppressed by only one
power of $\sin 2\theta_{13}$ (see eqs.~(\ref{Ps}) and (\ref{Pemudelta}))
and the corrections quadratic in $\dm{21}$ (CP-conserving) are not 
suppressed by $\theta_{13}$ at all. Indeed, at sufficiently small $L$ 
we have $\sin (\Delta \bar{E}_{m} L/2) \cong \Delta \bar{E}_{m} L/2$,
$\sin (\bar{\kappa} + \Delta \bar{E}_{m} L/2) \cong
(\bar{\kappa} + \Delta \bar{E}_{m} L/2)$, etc., and as can be shown 
\cite{SPtbp}, the leading CP-conserving and CP-violating 
$\dm{21}-$corrections containing the factor
$\cos \theta'_{12}$ in eq.~(\ref{Pemudelta}) reduce essentially to
their vacuum oscillation form:
\begin{equation}
\Delta P^{3\nu}_{E}(\nu_{e} \rightarrow \nu_{\mu};~\delta)
\cong \quad
8J^{V}_{CP}~\left ( {L\over 2}{{\Delta m^2_{21}}\over {2E}} \right ) 
~\left ( {L\over 2}{{\Delta m^2_{31}}\over {2E}} \right ) \quad  
\left [ \cot \delta ~-~ {L\over 2}{{\Delta m^2_{31}}\over {2E}}~
\right ]
\end{equation}
where 
\begin{equation}
J^{V}_{CP} = {1\over 8} \sin \delta ~c_{13} \sin 2\theta_{12}
\sin 2\theta_{13} \sin 2\theta_{23}
\end{equation}
is the CP-violation rephasing 
invariant of the lepton mixing matrix \cite{KP3nu88}. 
   
Unlike the appearance channels, the disappearance channels
are dominated by the transitions to $\nu_\tau$ ($\bar\nu_\tau$)
and are therefore not suppressed by $\theta_{13}$ or $\dm{21}$, 
so that the $\theta_{13}$-suppressed $\dm{21}$ corrections are 
much less significant in this case. Figs.~\ref{fig:sun1} 
and~\ref{fig:sun2} show also that the range of the $\dm{21}$ 
corrections is essentially determined by $\theta_{13}$, but the 
precise size and sign of these corrections within this range
is unknown if $\delta$ is unconstraint. This can be seen also 
explicitely in figs.~\ref{fig:sun1} and~\ref{fig:sun2} where 
the $\dm{21}$ corrections show up as oscillations around the 
leading $\dm{31}$ contribution to the rates (dashed line),
whose initial phase depends on $\delta$. The $\dm{21}$ effects 
represent consequently in the present LMA scenario in a high 
statistics long-but-not-too-long baseline measurement of $\theta_{13}$
an important source of systematic error, unless $\delta$ is 
measured~\cite{DFLR,romanino}. On the other hand, $\sin\delta$ 
could be measured or constrained in this scenario by comparing 
the rates in the two CP-conjugated channels $\reu$ and $\reub$ 
if very high intensity sources and large detectors will become 
available~\cite{romanino}. By comparing e.g. the left and right 
plots of in fig.~\ref{fig:sun2} one can see that, as it follows 
from eq.~(\ref{Pemudelta}), the $\dm{21}$ correction has the same 
sign in the two channels for the two CP-conserving values 
$\delta=0,\;\pi$, when $\sin\delta=0$, but has opposite sign for the
two CP-violating values $\delta=\pi/2,\;3\pi/2$, i.e. when
$\sin\delta=\pm1$. 

Note, however, that a measurement of $\delta$ based on a comparison of
CP-conjugated rates at a single value of $L$ could not be enough in
order to keep the $\dm{21}$ effects under control. Suppose, for example,
that $\delta =0$ or $\delta=\pi$. Then the $\dm{21}$ effects would be
the same in the two channels and an ideal experiment looking for
CP-violation would measure $\sin\delta = 0$. The value of $\delta$ 
would, however, still be undetermined for the simple reason that both
$\delta = 0$ and $\delta = \pi$ give $\sin\delta=0$. As a consequence,
it would be still unknown whether the $\dm{21}$ corrections in both
channels add ($\delta=0$) or subtract ($\delta=\pi$) to the leading
$\dm{31}$ contribution to the rates and this ambiguity would translate,
e.g.,\ in an uncertainty on a measurement of $\theta_{13}$. Such an
ambiguity could be resolved by comparing rates measured at different 
distances $L$. Fig.~\ref{fig:sun2} for $\sin^22\theta_{13}=0.01$ 
shows in the $\reu$ channel that the change in rates between 
$L=3000\,\text{km}$ and $L=700\,\text{km}$ allows to discriminate 
between $\delta=0$ and $\delta=\pi$. Fig.~\ref{fig:sun1} shows that 
the different $L$ dependence is also significant for 
$\sin^22\theta_{13}=0.1$, allowing also to distinguish between the 
two possibilities $\delta=0$ and $\delta=\pi$. If CP-violation were
maximal, $|\sin\delta|=1$, the comparison of the rates at different
baselines would be less significant but still helpful.
Figs.~\ref{fig:sun1} and~\ref{fig:sun2} show finally also that 
the $\dm{21}$ effects become smaller when $L$ is increased.

%%%%%%%%%%%%%%%%%%%%%%%%%%%%%%%%%%%%%%%%%%%%%%%%%%%%%%%%%%%%%%%%%%%%%
%%%%    SECTION: matter effects in spectrum                      %%%%
%%%%%%%%%%%%%%%%%%%%%%%%%%%%%%%%%%%%%%%%%%%%%%%%%%%%%%%%%%%%%%%%%%%%%

\section{Matter Effects in the Energy Spectrum.}
\label{sec:spectrum}

Motivated by the small differential event rates, we discussed so far
only the influence of matter effects on the total rates of wrong
sign muon events. We demonstrated in the previous chapters that
statistical significant deviations from the total event rates in
vacuum represent already a good test of the MSW theory. A
significant test would however be also given by a detailed measurement
of MSW effects in the neutrino energy spectrum, which is modified in a
very characteristic way.  The $\ruu$ and $\ruub$ disappearance 
channels are again dominated by transitions to $\nu_\tau$ and 
$\bar\nu_{\tau}$ while $\nu_e$ and $\bar\nu_{e}$ transitions are 
only small corrections. These channels are therefore mostly 
insensitive to matter effects in the differential event rate 
spectrum and will not be discussed further.

\begin{figure}[ht!]
\begin{center}
\epsfig{file=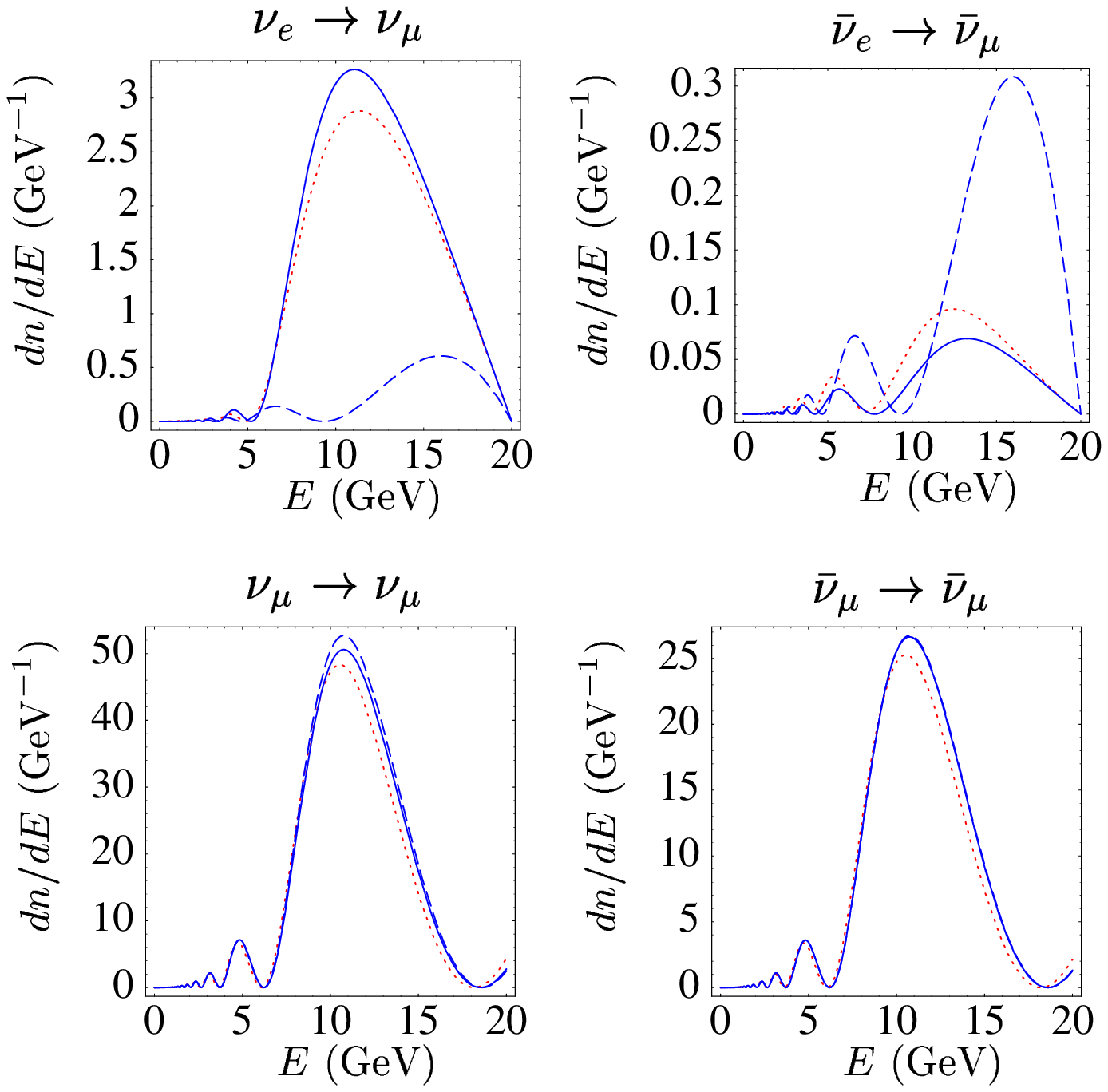,width=0.80\textwidth}
\end{center}
\mycaption{Modifications in the differential event rate 
spectrum (events per $\GeV$) due to matter effects for 
$E_\mu=20\GeV$. The solid lines correspond to oscillations
taking place in the earth mantle, while the dashed lines are 
for oscillations in vacuum, with $\Delta m^2_{21}=0$ in 
both cases. The ``enhancement'', the ``broadening'' 
and the ``shift''  of the first oscillation maximum in the 
$\nu_e \rightarrow \nu_{\mu}$ channel as $E$ decreases, 
caused by the MSW effect, is clearly seen. The dotted lines 
show for comparison an example of $\Delta m^2_{21}$ 
corrections for $\Delta m^2_{21} = 10^{-4}~\mathrm{eV}^2$, 
$\delta = 0$ and  $\sin^22\theta_{12} = 0.8$. The assumed 
parameters are in all cases $N_\mu\NKT\epsilon=10^{21}$, 
$L=6596\,\text{km}$ (CERN-MINOS),
$\Delta m^2_{31} = 3.5 x 10^{-3}~\mathrm{eV}^2$
and $\sin^2 2\theta_{13}=0.01$.}
\label{fig:spectrum20}
\end{figure}
\begin{figure}[ht!]
\begin{center}
\epsfig{file=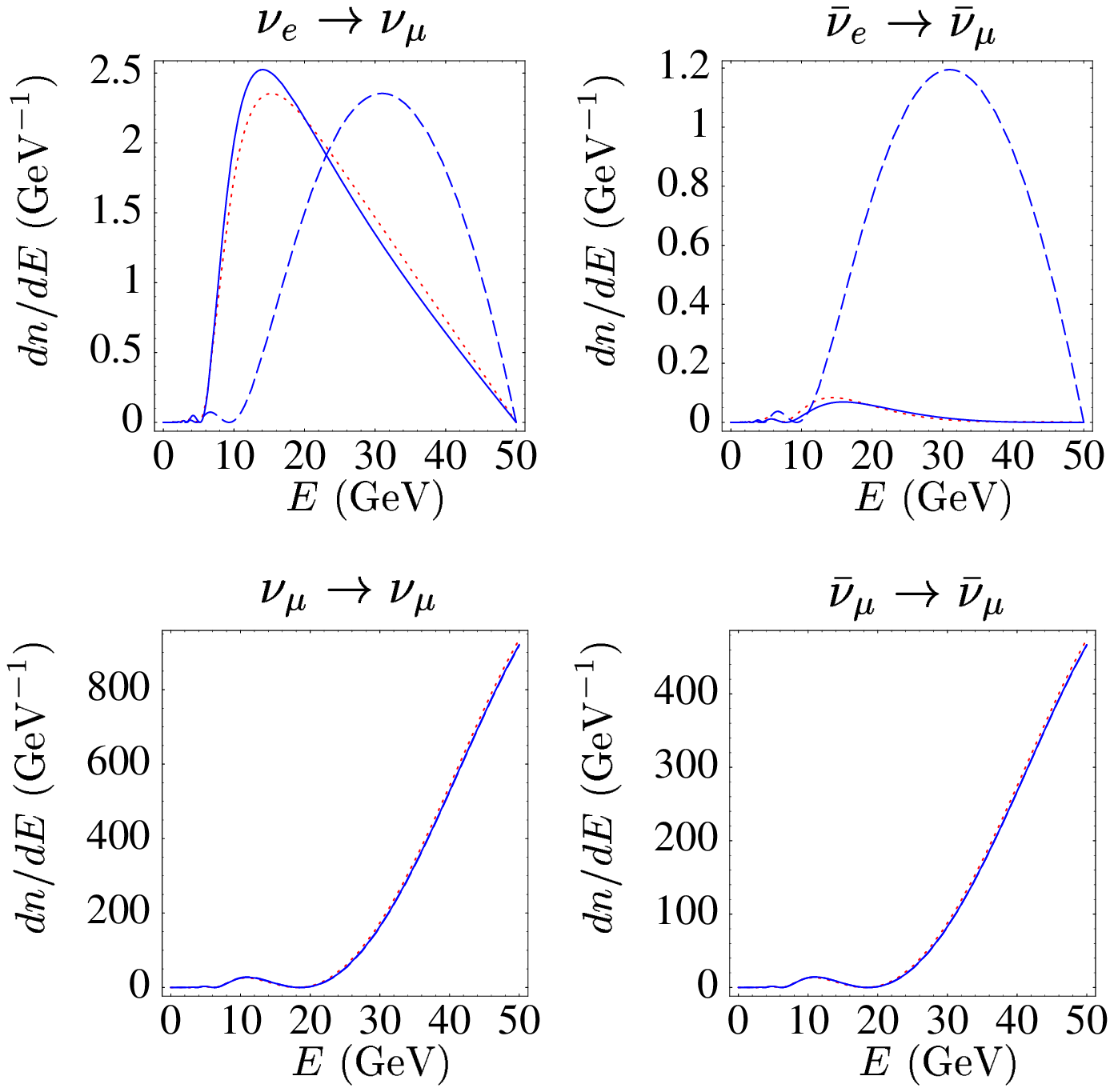,width=0.80\textwidth}
\end{center}
\mycaption{Same as in fig.~\ref{fig:spectrum20} but for 
$E_\mu=50\GeV$. For details see text.
}
\label{fig:spectrum50}
\end{figure}
To understand the effects in the appearance spectrum of $\reu$ 
and $\reub$ we us again the approximation $\Delta m^2_{21}=0$. 
Matter effects have no influence on the angle $\theta_{23}$ in 
this case, whereas they modify the mixing due to $\theta_{13}$ 
significantly. One obtains thus for the $\reu$ and $\reub$ 
appearance channels the usual two flavour picture where 
$\sin2\theta_{13}\rightarrow\sin2\theta_{13}^{\text{m}}=
\sin2\theta_{13}/C_\pm$, and where $C_+$ corresponds to neutrinos, 
$C_-$ to antineutrinos. The enhancement of $\theta^m_{13}$ is 
maximal in the neutrino channel when the neutrino energy $E$ 
coincides with the MSW resonance energy $\eres$ defined via:
\begin{equation}
\label{eres}
\frac{2\eres V}{\dm{31}}=\cos 2\theta_{13}~.
\end{equation}
Note however that the maximum of the event spectrum in the 
$\reu$ channel does in general not coincide with $\eres$
since the MSW oscillation probabilities are folded with 
the fluxes and cross sections. The maximum of the event 
spectrum is thus determined by the maximization of
\begin{equation}
\label{max}
\left(\frac{\sin(\Delta_{31}
C_\pm)}{C_\pm}\right)^2f_{\nu_e\nu_\mu}(E/E_\mu)~,
\end{equation}
but the resulting maximum is for the muon energies under 
consideration still around $\eres$. The offset 
depends in a rough approximation on the difference 
between $\eres$ and the maximum of the flux which lies 
roughly at an energy of the order $E_\mu$.

This has to be compared with the vacuum case where the 
oscillation probabilities are also folded with fluxes and 
cross sections and where the maxima of the event 
rates are also not precisely at the maxima 
of the oscillation probabilities. The event rate spectrum 
is thus due to this folding in both cases with and without 
matter a rather complicated function of 
$E_\nu$ which depends in a non-trivial way on $L$,
$E_\mu$ and $\dm{31}$. Nevertheless for given $L$, $E_\mu$ 
and for given $\dm{31}$, $\theta_{23}$, $\theta_{13}$ 
measured with a suitable long baseline experiment, 
one can predict the shape  of the differential event rate 
spectrum in all channels and compare it with the spectrum 
of oscillations unaffected by matter. 
This results in a very good opportunity to detect 
specific details of the MSW effect which arise when the 
oscillation parameters are chosen such that the first 
maximum of vacuum oscillation coincides roughly with 
$\eres$. The point is that the MSW effect changes the 
probabilities compared to vacuum in three genuine ways: 
The first maximum of the oscillation probability as the 
energy decreases is enhanced, its width is broadened and 
its center is shifted to lower energies. Similarly
one has an ``anti--MSW effect'' in the antineutrino
appearance channel which implies for the first oscillation
maximum a reduction in height, again a broadening and a 
shift to lower energies. 

These ``genuine MSW effects'' are demonstrated in
fig.~\ref{fig:spectrum20} (where $E_\mu=20\GeV$) and
fig.~\ref{fig:spectrum50} (with $E_\mu=50\GeV$) showing the
modifications in the energy spectrum (events per $\GeV$) due to 
matter effects.  The solid lines are in matter while the dashed 
lines are without matter and the assumed parameters are
$N_\mu\NKT\epsilon=10^{21}$ as before, $L=6596\,\text{km}$ and 
$\sin^2 2\theta_{13}=0.01$.  Fig.~\ref{fig:spectrum20} shows 
already all effects due to the MSW mechanism, namely the broadening 
(the last oscillation in matter covers two oscillations in vacuum), 
the shift (the maximum in matter lies almost in the minimum in 
vacuum) and the enhancement or suppression compared to vacuum.
Figs.~\ref{fig:spectrum20} and~\ref{fig:spectrum50} also include an
example of $\dm{21}$ corrections to the spectrum in the LMA scenario
(dotted lines). Size and sign of the corrections depend on the value 
of the phase $\delta$. The dotted line assumes $\delta = 0$ and
moreover $\dm{21}=10^{-4}\eV^2$ and $\sin^22\theta_{12}=0.8$. 
As already observed in the previous section, the $\dm{21}$ corrections 
are relatively small for the very long baselines considered here.

It is interesting to look at the modifications when the muon energy
becomes higher, e.g. for $E_\mu=50\GeV$ as shown in
fig.~\ref{fig:spectrum50}. The point is that the beam energy is
already rather far away from the MSW resonance energy and the
importance of the weight function $f_{\nu_e\nu_\mu}$ (and of the
$E^3_\nu$ scaling of unoscillated events) in the determination of the
shape of the spectrum increases. Thus the genuine broadening,
shift and enhancement/suppression effects become harder to
distinguish.  For the $\reu$ channel the effect could be hard to
distinguish from uncertainties (with low statistics) in the spectrum.
This brings up the general issue that one has to have enough
statistics for such an analysis. In order to have a chance to see such
effects one has to be lucky and $\theta_{13}$ should be at the upper
experimental limit (see scaling laws). Otherwise $N_\mu\NKT\epsilon$
must be increased correspondingly which implies a more intense muon
source, a larger detector or both.  Although a more quantitative
analysis of the significance of effects in the differential neutrino
event rate spectrum is beyond the scope of this paper, an analysis of
the differential event rate spectrum would clearly provide extremely
valuable additional information which would allow to test some of the
characteristic features of the MSW mechanism.

%%%%%%%%%%%%%%%%%%%%%%%%%%%%%%%%%%%%%%%%%%%%%%%%%%%%%%%%%%%%%%%%%%%%%
%%%%    SECTION: conclusions                                     %%%%
%%%%%%%%%%%%%%%%%%%%%%%%%%%%%%%%%%%%%%%%%%%%%%%%%%%%%%%%%%%%%%%%%%%%%

\section{Conclusions \label{sec:conc}}

Assuming three-neutrino mixing we studied in this paper the 
possibility to test the MSW effect in terrestrial very long baseline  
neutrino oscillation experiments which become possible with 
neutrino factories. Such direct tests are important since the 
MSW mechanism is widely used in different scenarios of neutrino 
physics and astrophysics. The correct analysis and interpretation 
of the data from terrestrial very long baseline neutrino oscillation 
experiments, $L \gtap 1000$ km, is in fact impossible without 
a proper treatment of matter effects. The latter is also crucial
for the searches of CP-violation in neutrino oscillations
generated by the lepton mixing matrix, since matter effects create 
an asymmetry between the two CP-conjugated appearance channels. 

Studies of the $\reu$ and $\reub$ oscillations are by far most 
promising for the detection of the matter effects since the 
corresponding total event rates are affected in a drastic way
by these effects.
We considered for the present study neutrino trajectories 
through the earth which cross the mantle, but do not pass 
through the earth core, which corresponds to neutrino path 
lengths $L \ltap 10600$~km. Using analytic expressions for the 
three neutrino oscillation probabilities in matter in the 
constant average density and small $\Delta m^2_{21}$ approximations
and including fluxes, cross sections and detection efficiencies 
allows to describe the relevant event rates analytically.
This permitted a full analytic understanding of our numerical 
results.

By considering the asymmetry between the $\reu$ and $\reub$ 
induced wrong sign muon event rates, we studied the statistical 
significance of the observation of matter effects as a function 
of the neutrino oscillation parameters $\theta_{13}$ and 
$\Delta m^2_{31}$ as well as its dependence on the experimental 
conditions via the parent muon beam energy $E_{\mu}$, the path 
length $L$ and the product of useful muons, detector size and 
efficiency. The scaling of rates, statistical significances 
and sensitivities with the relevant mixing angles, in 
particular, with $\theta_{13}$,  the intensity of the muon 
source and with the detector size and efficiency have been 
given, so that the results for any value of those parameters 
can easily be obtained. The sign of the asymmetry  depends   
on whether the two closest neutrino mass eigenstates are lighter 
($\dm{31}>0$) or heavier ($\dm{31}<0$) than the third one,
thus providing a way of determining which of these two 
possibilities is realized. 
Figs.~\ref{fig:CL20} and~\ref{fig:CL50} show the
conservative ranges of $\sin^22\theta_{13}$ where that 
determination would be significant at a given confidence level 
from the statistical point of view as function of the baseline. 

We analyzed, in particular, the statistical significance of
matter effects as a function of $E_{\mu}$ and $L$. The most
important requirement regarding the muon energy is that for
given $\Delta m^2_{31}$ it has to be greater than the MSW
resonance energy. For, e.g.,
$\Delta m^2_{31} \leq 6.0~(8.0)\times 10^{-3}$~eV$^2$,
this implies $E_{\mu} \gtap 20~(30)$~GeV. Using a simple 
``total rates based'' method and assuming that $\theta_{13}$ 
is known, any $L$ from the interval $(4.0 - 10.0)\times 10^{3}$~km 
gives relatively good sensitivity to matter effects if 
$\Delta m^2_{31} = 3.5\times 10^{-3}\eV^2$, 
$\sin^22\theta_{13} = 0.01$ and $E_{\mu} = 20$ GeV, with the 
sensitivity increasing with $L$. For $E_{\mu} = 50$ GeV, the 
sensitivity varies very little for 
$L \cong (4.0 - 10.0)\times 10^{3}$~km. Let us note that if
$\Delta m^2_{31} \simeq 10^{-3}$~eV$^2$ and 
$\sin^22\theta_{13} \ltap 0.01$, then establishing the matter effects 
(or obtaining a stringent upper limit on $\sin^22\theta_{13}$) 
would require $E_{\mu} \simeq (40 - 50)$~GeV.
As our results show, the optimal value of 
$E_{\mu}$ depends on the precise value of 
$\Delta m^2_{31}$.

Our analysis shows that a higher sensitivity to the MSW effect 
in the case of relatively small values of $\sin^22\theta_{13}$ 
is achieved at very large $L$. We showed that this conclusion 
holds also when subleading $\Delta m^2_{21}$ effects are 
included. These effects can be significant in the case of the 
LMA~MSW solution with 
$\Delta m^2_{21} \simeq (0.5 - 2.0)\times 10^{-4}$~eV$^2$.
At very large $L$, the indicated $\Delta m^2_{21}$-induced 
effects are essentially 
negligible and the results obtained in the limit of 
$\Delta m^2_{21} = 0$ are therefore sufficiently accurate. 
This is valid even in the case of the LMA~MSW solution with
$\Delta m^2_{21} \simeq (0.5 - 2.0)\times 10^{-4}$~eV$^2$.
For shorter baselines, $\dm{21}$ effects are non-negligible 
and a determination of the CP-violating phase $\delta$ would 
be necessary in order to know their precise magnitude. 
We have found that the $L$ dependence of the $\dm{21}$ effects 
offers the possibility to determine the CP-phase $\delta$, 
especially when $\sin\delta$ is small. 
For $\sin^22\theta_{13} = 0.01$, for instance, the event 
rates due to the $\reu$ and $\reub$ transitions change 
considerably when $L$ changes from $\sim 700$~km to 
$\sim 3000$~km. Thus, a measurement of these rates, e.g.,
at the indicated two distances could allow to determine 
the value of $\delta$ with a certain precision.
Finally we discussed the matter effects in the differential
event rate spectrum as a function of the neutrino energy, 
and showed that they lead to very characteristic distortions.
The observation of these distortions would allow
very detailed tests of the MSW theory.

To conclude, our study shows that the predictions of the 
MSW theory can be tested in a statistically reliable way 
in a large region of the corresponding parameter space by 
a simple analysis of the total event rates in a very long 
baseline neutrino oscillation experiment. This can allow 
to determine the sign of $\Delta m^2_{31}$ as well, as was 
recently noticed also in \cite{BARGER99b}. Not seeing the 
matter effects would lead to impressive upper limits on the 
mixing angle $\theta_{13}$ down to 
$\sin^22\theta_{13}\simeq 10^{-4}$ or even better.

%%%%%%%%%%%%%%%%%%%%%%%%%%%%%%%%%%%%%%%%%%%%%%%%%%%%%%%%%%%%%%%%%%%%
%%%%                      Acknowledgments                      %%%%
%%%%%%%%%%%%%%%%%%%%%%%%%%%%%%%%%%%%%%%%%%%%%%%%%%%%%%%%%%%%%%%%%%%%

\vspace*{7mm}
Acknowledgments: A.R. and S.P. wish to thank the Institute T30d 
at the Physics Department of the Technical University of Munich 
for warm hospitality.

%%%%%%%%%%%%%%%%%%%%%%%%%%%%%%%%%%%%%%%%%%%%%%%%%%%%%%%%%%%%%%%%%%%%%
%%%%                       References                            %%%%
%%%%%%%%%%%%%%%%%%%%%%%%%%%%%%%%%%%%%%%%%%%%%%%%%%%%%%%%%%%%%%%%%%%%%

\newpage

\bibliographystyle{phaip}

\begin{thebibliography}{1}

\bibitem{GEER98} S.~Geer, 
Phys. Rev. {\bf D57} (1998) 6989, (E) {\sl ibid.} {\bf D59} (1999) 039903.

\bibitem{gavela-c}
A. De Rujula, M.B. Gavela and P. Hernandez, 
Nucl. Phys. {\bf B547} (1999) 21.

\bibitem{BARGER99}
V.~Barger, S.~Geer and K.~Whisnant, e-Print Archive: hep-ph/9906487.

\bibitem{DFLR} K. Dick, M. Freund, M. Lindner and A. Romanino,
Nucl. Phys. {\bf B562} (1999) 29.

\bibitem{BCR9905240} M. Campanelli, A. Bueno and A. Rubbia, 
e-Print Archive: hep-ph/9905240. 

\bibitem{gavela-b}
A. Donini, M.B. Gavela, P. Hernandez and S. Rigolin,
e-Print Archive: hep-ph/9909254. 

\bibitem{romanino}
A.~Romanino, e-Print Archive: hep-ph/9909425, 
to be published in Nucl. Phys {\bf B}.

\bibitem{BARGER99b} V. Barger, S. Geer, R. Raja and K. Whisnant,
e-Print Archive: hep-ph/9911524.

\bibitem{lyon}
Nufact '99 Workshop, July 5--9th, Lyon. See e.g.\ the Summary of
Detector/Neutrino Beam Parameters by D. Harris, 
{http://lyopsr.in2p3.fr/nufact99/talks/harrisconc.ps.gz}.

\bibitem{Lipari} P. Lipari, e-Print Archive: hep-ph/9903481.

\bibitem{Mohan} M. Narayan and S.U. Sankar,
Phys. Rev. {\bf D61} (2000) 013003.
  
\bibitem{Stacey77} F.D. Stacey,
{\it Physics of the Earth}, 2$^{nd}$ edition,
John Wiley and Sons, New York, 1977.

\bibitem{PREM81} A.D. Dziewonski and D.L. Anderson,
Phys. Earth Planet. Interiors {\bf 25} (1981) 297.

\bibitem{SP1} S.T. Petcov, Phys. Lett. {\bf B434} (1998) 321, 
(E) {\sl ibid.} {\bf B444} (1998) 584. 

\bibitem{MLiuSP96} M. Maris and S.T. Petcov, unpublished.

\bibitem{MartinTommy} M. Freund and T. Ohlsson, 
e-Print Archive: hep-ph/9909501.  

\bibitem{Shrock99} I. Mocioiu and R. Shrock, e-Print Archive: hep-ph/9910554.

\bibitem{MannP77} A.K. Mann and H. Primakoff,
Phys. Rev. {\bf D15} (1977) 655;\\
V.K. Ermilova, V.A. Tsarev V.A. Chechin,
Zh.Eksp.Teor.Fiz. (Pis'ma) {\bf 43} (1986) 3531;\\ 
P.I. Kratsev, Il Nuovo Cimento {\bf 103 A} (1990) 361. 

\bibitem{magdetector}
See e.g.\ ``Detectors working group + Neutrinos Beam Parameters Working
Group'' in the proceeding of the Nufact '99 Workshop, July 5--9th, Lyon,
or the ``Oscillation Detectors'' working group web page at
http://www.cern.ch/muonstoragerings/.

\bibitem{solardm2}
G.L. Fogli, E. Lisi, D. Montanino and A. Palazzo,
e-Print Archive: hep-ph/9912231 and 9910387;
M. Nakahata, in {\it TAUP'99}, VIth Internationa Workshop
on Topics in Astroparticle and Underground Physics,
September 6 - 10, 1999, Paris, France, to appear
(transparanceis available at http://taup99.in2p3.fr/TAUP99/).

\bibitem{atmdm2}
K. Scholberg, for the Super-Kamiokande Collab., 
e-Print Archive: hep-ex/9905016.

\bibitem{ADeR80} A. De Rujula, M. Lusignoli, L. Maiani, 
S.T.~Petcov and R.~Petronzio, Nucl. Phys. {\bf B168} (1980) 54.

\bibitem{BPet87} S.M. Bilenky and S.T. Petcov,
Rev. Mod. Phys. {\bf 59} (1987) 671.

\bibitem{CHOOZ99} M. Apollonio et al. (CHOOZ collaboration),
e-Print Archive: hep-ex/9907037.

\bibitem{CSL87} C.-S. Lim, Report BNL 52079 (1987).

\bibitem{3nuSP88} S.T. Petcov, Phys. Lett. {\bf B214}, (1988) 259.

\bibitem{3nu} T.K. Kuo and J. Pantaleone, 
Phys. Rev. Lett. {\bf 57} (1986) 1805;\\ 
S.T. Petcov and S. Toshev, Phys. Lett. {\bf B187} (1987) 120;\\
A. Yu. Smirnov, Yad. Fiz. {\bf 46} (1987) 1152;\\ 
see also:
S. Toshev, Phys. Lett. {\bf B185} (1987) 177, 
(E) {\sl ibid.} {B\bf 192} (1987) 478;\\ 
C.W. Kim, S. Nussinov and W.K. Sze, Phys. Lett. {\bf B184} (1987) 403;\\ 
A. Baldini and G.F. Giudice, Phys. Lett. {\bf B186} (1987)  211.

\bibitem{SPWIN99} S.T. Petcov, Invited talk given at the 
Int. Workshop on Weak Interaction and Neutrinos,
January 25 - 30, 1999, Cape Town, South Africa
(to be published in the Proceedings of the Workshop),
e-Print Archive: hep-ph/9907216.

\bibitem{OY99} O. Yasuda, e-Print Archive: hep-ph/9910428. 

\bibitem{BHP80}  S.M. Bilenky, J. Hosek and S.T. Petcov, 
Phys. Lett. {\bf B94} (1980) 495;\\
P. Langacker, S.T. Petcov, G. Steigman and S.~Toshev, 
Nucl. Phys. B{\bf 282} (1987) 589.

\bibitem{s5398} M. V. Chizhov, M. Maris and S. T. Petcov, 
e-Print Archive: hep-ph/9810501.

\bibitem{KP3nu88} P.I. Krastev and S.T. Petcov,
Phys. Lett. {\bf B205} (1988) 84.

\bibitem{ADLS99} E.K. Akhmedov, A. Dighe, P. Lipari and A.Yu. Smirnov, 
Nucl. Phys. {\bf B542} (1999) 3.      

\bibitem{SPtbp} S.T. Petcov, in preparation.

\bibitem{CDG}
A.~Cervera~Villanueva, F.~Dydak and J.~{G\'omez}-Cadenas,
\newblock {N}ufact '99 {W}orkshop, July 5--9th, Lyon,
{http://lyopsr.in2p3.fr/nufact99/talks/cervera.ps.gz}.

\bibitem{PDG99}
C. Caso et al., The European Physical Journal {\bf C3} (1998) 1,
and 1999 off-year partial update for the 2000 edition available on 
the Particle Data Group WWW pages (URL: http://pdg.lbl.gov/). 

\bibitem{concha} M.C. Gonzalez-Garcia, P.C. de Holanda, 
C. Pena-Garay and J.W.F.~Valle, e-Print Archive: hep-ph/9906469. 

\bibitem{KuoP87} P.K. Kuo and J. Pantaleone, 
Phys. Lett. {\bf B198} (1987) 406.

\bibitem{sato1}
M. Koike and J. Sato, e-Print Archive: hep-ph/9909469 and
hep-ph/9911258.

\end{thebibliography}

\end{document}